\documentclass[pre,amsmath,amssymb,singlecolumn,superscriptaddress]{revtex4}
\usepackage{amsmath,amssymb}
\usepackage[usenames]{color}
\usepackage{amssymb}
\usepackage{grffile}
\usepackage{amsmath, amstext, amssymb, amsfonts, amsxtra}
\usepackage{textcomp}
\usepackage{xspace}
\usepackage{float}
\usepackage{amsmath}
\usepackage{amssymb}
\usepackage{mathtools}
\usepackage{braket}
\usepackage{subfig}
\usepackage{graphicx}
\usepackage{listings}
\usepackage{color}
\definecolor{mygreen}{rgb}{0,0.6,0}
\definecolor{mygray}{rgb}{0.5,0.5,0.5}
\definecolor{mymauve}{rgb}{0.58,0,0.82}

\usepackage{amssymb}
\usepackage{latexsym}

\usepackage{url}
\usepackage{xcolor}
\definecolor{newcolor}{rgb}{.8,.349,.1}

\newcommand{\vvec}{\text{vec}}
\newcommand{\trans}{^{\text{T}}}
\DeclareMathOperator{\tr}{tr}

\newcommand{\usp}{Instituto de F\'isica, Universidade de S\~ao Paulo, 05314-970, S\~ao Paulo, S\~ao Paulo, Brasil}
\newcommand{\sutd}{Science, Mathematics and Technology Cluster, Singapore University of Technology and Design, 8 Somapah Road, 487372 Singapore}
\newcommand{\epd}{Engineering Product Development Pillar, Singapore University of Technology and Design, 8 Somapah Road, 487372 Singapore}

\begin{document}

\title{Analysis of a density matrix renormalization group approach for transport in open quantum systems}

\author{Heitor P. Casagrande}
\email{heitor\_peres@mymail.sutd.edu.sg}
\affiliation{\sutd}
\affiliation{\usp} 

\author{Dario Poletti}
\email{dario\_poletti@sutd.edu.sg}
\affiliation{\sutd}
\affiliation{\epd}

\author{Gabriel T. Landi}
\email{gtlandi@if.usp.br}
\affiliation{\usp}





\begin{abstract}
Understanding the intricate properties of one-dimensional quantum systems coupled to multiple reservoirs poses a challenge to both analytical approaches and simulation techniques. Fortunately, density matrix renormalization group-based tools, which have been widely used in the study of closed systems, have also been recently extended to the treatment of open systems. We present an implementation of such method based on state-of-the-art matrix product state (MPS) and tensor network methods, that produces accurate results for a variety of combinations of parameters. 
Unlike most approaches, which use the time-evolution to reach the steady-state, we focus on an algorithm that is time-independent and focuses on recasting the problem in exactly the same language as the standard Density Matrix Renormalization Group (DMRG) algorithm, initially put forward in \cite{LdLTechnique}.
Hence, it can be readily exported to any of the available DMRG platforms. 
We show that this implementation is suited for studying thermal transport in one-dimensional systems. 
As a case study, we focus on the XXZ quantum spin chain and benchmark our results by comparing the spin current and magnetization profiles with analytical results. We then explore beyond what can be computed analytically.
Our code is freely available on github at \cite{oDMRG}.
\end{abstract}

\maketitle

\section{Introduction}

Transport properties at the nanoscale may be significantly different from bulk materials, because of low-dimensionality, interactions and interference. For instance in quantum spin chains one can observe anomalous diffusion \citep{Landi2015, Marko2011, Znidaric2009}, negative differential conductance  and rectification \citep{LandiRectification, EmmanuelRectification, DarioEmmanuelRectificationXXZ}.

\begin{figure}
	\centering	
	\includegraphics[scale = 0.45]{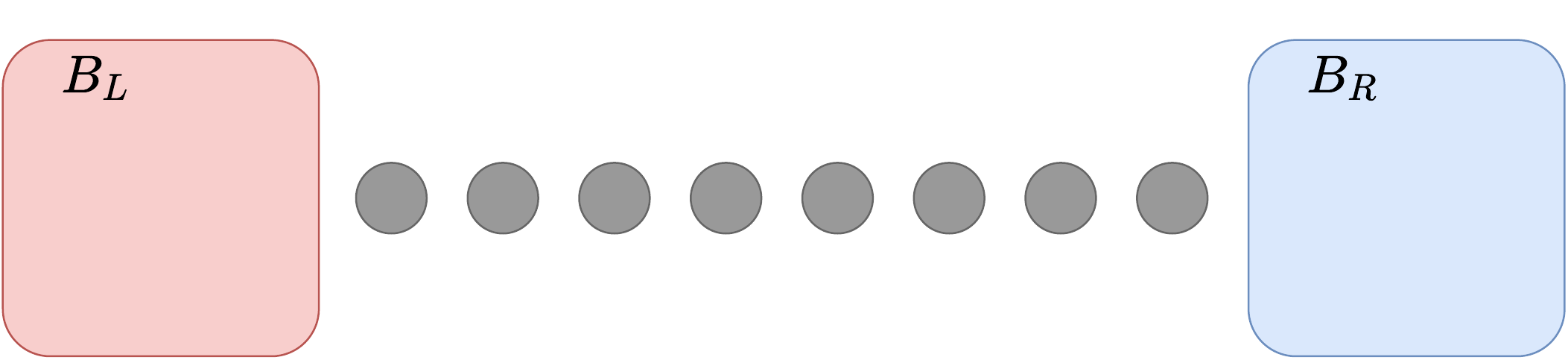}
	\caption{Diagrammatic representation of a one-dimensional quantum spin chain coupled to two thermal baths $B_L$ and $B_R$, one at each end. Here, each grey circle represent a spin.}
	\label{ExampleOQS}
\end{figure}

The study of many-body quantum systems, however, is in general very demanding. For instance, considering a pure state describing a chain of $N$ spin-$1/2$ particles, one needs to take into account a Hilbert space of size $2^N$. However, when aiming to study the transport properties of a system coupled to two different baths at its edges, as shown in Fig.\ref{ExampleOQS}, one needs also to find a way to model the effects of the bath, thus requiring to explore a space larger than $2^N$. One approach to study open quantum, i.e. quantum systems in contact with an environment, is that of the Gorini-Kossakowski-Sudarshan-Linbdlad (GKSL) master equation, which is a linear equation describing the evolution of the density matrix of the system \citep{LindbladOriginal, Gorini1976}. For this reason, to study transport in this framework, one needs to be able to explore a space of dimension $2^{2N}$. An exact study of systems of this type becomes quickly too difficulty, e.g. for systems with $L\approx 12$ (see \citep{GuoPoletti2018} for an exact diagonalization study with $L=14$, which was only possible by considering symmetries of a particular class of boundary driven problems). 

Fortunately, for the particular case of one-dimensional (1D) systems, a particular class of numerical methods has been developed starting from the seminal work \citep{FirstDMRG}, which outlined the density matrix renormalization group (DMRG) method that was later realized within the general framework of tensor networks \citep{SchollwockMPS}. Tensor networks are a particular form of variational ansatz to explore, at a polynomial cost, an otherwise exponentially large Hilbert space. The key advantage of this variational approach is that a large class of physically relevant ground states, e.g. Hamiltonians of 1D systems with finite range interactions, can be exactly described using tensor networks \citep{SchollwockMPS}. The key is that the bipartite entanglement entropy of the system should not grow linearly with the system size (a.k.a. volume law). 

Tensor network methods are known to be useful also in the description of open quantum systems. 
In this case there is no analytical proof that the method will accurately describe  the open quantum system.
However, as we will see later in this manuscript, the method can be very accurate for many physically relevant scenarios. 
Various approaches have been put forward to study many-body open quantum systems with tensor networks. 
A review which focuses on ensemble trajectories of stochastic wavefunctions \citep{DalibardCastin1993},  instead of evolving the density matrix, can be found in \citep{Daley2014}.
And a comparison of the trajectory method versus evolving directly the density matrix can be found here \citep{Lauchli, Kollath}. A short review of numerical methods to study many-body open quantum systems can be found in \citep{WeimerOrus}. A particularly interesting approach was put forward in \citep{LdLTechnique}, in which the evaluation of the steady-state of a system was mapped into that of computing the ground state of an effective Hamiltonian with long range interactions. In this work we evaluate the performance of this approach when applied to boundary driven systems as the one shown in Fig.\ref{ExampleOQS}. 

To make the paper self-contained, we first provide in Sec.\ref{sec:model} a detailed description of the model studied, with all the relevant equations and the different transport properties that can emerge.
We follow this with a short description of a tensor network algorithm to compute the ground state for closed systems in Sec.~\ref{sec:TN}. 
Then, in Sec.\ref{sec:HilbertSpace} we discuss in detail how one can map an open quantum system problem to a form conducive for tensor networks calculations.
Next we describe how one can map the problem of finding the steady state of a many-body open quantum system to that of computing the ground state of an effective Hamiltonian in Sec.\ref{sec:implementation}, where we also provide all details of an implementation using the ITensor library \cite{ITensor}.
A series of numerical analyses benchmarking our code, comparing numerical and analytical results, and exploring the physics beyond what can be addressed analytically, is then discussed in Sec.~\ref{sec:results}. Conclusions are given in Sec.~\ref{sec:conclusions}.

\section{A boundary driven spin chain}\label{sec:model} 

One of the most widely studied examples of a open quantum spin chain is the XXZ model coupled to two local GKSL baths. 
The Hamiltonian for a 1D chain of $N$ sites is given by 
\begin{equation}\label{H}
H = \sum\limits_{i=1}^{N-1}  J_i \bigg( \sigma_i^x \sigma_{i+1}^x + \sigma_i^y \sigma_{i+1}^y + \Delta_i \sigma_i^z \sigma_{i+1}^z \bigg) + \sum\limits_{i=1}^N h_i \sigma_i^z, 
\end{equation}
where $\sigma_i^\alpha$ are the Pauli matrices and $J_i, \Delta_i, h_i$ are parameters indicating respectively the tunneling between nearest sites, the anisotropy and a local magnetic field.   
In addition to the Hamiltonian dynamics, the system is also coupled to two baths at sites $1$ and $N$, as described by a GKSL master equation \citep{MasterEq}. 
The evolution of the system's density matrix $\rho$ will then be given by 
\begin{equation}\label{M}
\frac{d \rho}{d t} = \mathcal{L}(\rho) := -i [H,\rho] + D_1(\rho) + D_N(\rho),
\end{equation}
where 
\begin{equation}\label{D}
D_i(\rho) = \gamma_i f_i \mathcal{D}[\sigma_i^-](\rho) + \gamma_i (1-f_i) \mathcal{D}[\sigma_i^+](\rho), \qquad \quad i = 1,N,
\end{equation}
with $\mathcal{D}[L](\rho) = L \rho L^\dagger - \frac{1}{2} \{ L^\dagger L, \rho\}$. 
Here $\gamma_i > 0$ represent the coupling strength to bath $i$ and $f_i \in [0,1]$ represent the imbalance between the baths.   
After a sufficient time has elapsed, the evolution of Eq.~(\ref{M}) will eventually reach a non-equilibrium steady-state (NESS) defined by 
\begin{equation}\label{NESS}
\mathcal{L}(\rho_\text{ness}) = 0. 
\end{equation} 
In the vast majority of cases, this steady-state is also unique. 
Note also that, albeit a steady-state, the system will not be in equilibrium since there will be, in general, a steady current flow from one bath to the other. 

The model described by Eqs.~(\ref{H})-(\ref{D}) presents remarkably rich physics. 
The most relevant observables to analyze are the local currents from site $i$ to $i+1$
\begin{equation}\label{current}
\mathcal{J}_i = 2 J_i (\sigma_x^i \sigma_y^{i+1} - \sigma_y^{i} \sigma_x^{i+1}), 
\end{equation}
and the local magnetization $\sigma_z^i$. 
In the NESS, current conservation implies that $\langle \mathcal{J}_i \rangle$ will be independent of $i$ (the current from $i-1 \to i$ is the same as that from $i\to i+1$).
The physics is then characterized by the different transport properties of $\langle \mathcal{J}_i \rangle$. 
For large sizes, one usually has the scaling
\begin{equation}
\langle \mathcal{J}_i \rangle \sim \frac{1}{L^\alpha},
\end{equation}
where $\alpha >0$ is an exponent characterizing the type of transport: ballistic for $\alpha = 0$, diffusive for $\alpha = 1$, superdiffusive for $\alpha \in [0,1]$, subdiffusive for $\alpha > 1$, and insulating when $\alpha\rightarrow \infty$.

\section{\label{sec:TN}Review of Tensor Network methods in closed quantum systems} 

\begin{figure}
	\centering	
	\includegraphics[trim= 0 440 0 200, clip, scale=0.35]{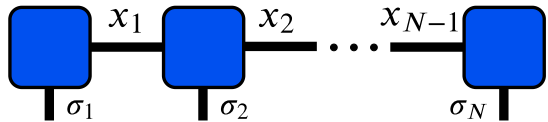}
	\caption{Tensor network representation [Eq.~(\ref{MPS})] of a high-rank tensor as a contraction of lower rank tensors. }
	\label{PsiMPS}
\end{figure}

The basic idea behind tensor networks is to decompose a high-rank tensor into a controlled product of lower rank tensors. 
Consider a generic rank-$N$ tensor $\psi_{\sigma_1\ldots \sigma_N}$. A tensor network decomposition has the form 
\begin{equation}\label{MPS}
\psi_{\sigma_1\ldots\sigma_N} = \sum\limits_{x_1,\ldots, x_{N-1}} A^{\sigma_1}_{x_1} A^{\sigma_2}_{x_1, x_2} ... A^{\sigma_N}_{x_{N-1}},
\end{equation}
which is shown diagrammatically in Fig.~\ref{PsiMPS}.
This kind of expansion is relevant because quantum states of multipartite systems are naturally represented as a high-rank tensor. 
For instance, the state of a spin chain with $N$ sites has the form 
\begin{equation}
\ket{\Psi} = \sum\limits_{\sigma_1, \ldots, \sigma_N} \psi_{\sigma_1,\ldots, \sigma_N} \ket{\sigma_1, \ldots, \sigma_N},
\label{PsiChainL}
\end{equation}
where $\sigma_i = \pm 1$ are the eigenvalues of $\sigma_i^z$. 
Of course, while a decomposition of the form~(\ref{MPS}) is always possible, it is not necessarily advantageous. 
The advantages ultimately come from approximations that can be obtained by restricting the dimension of the internal indices $x_i$, called the bond dimension.


The tensor network decomposition~(\ref{MPS}) is used as the starting point for a variety of algorithms. 
The most notable is the Density Matrix Renormalization Group (DMRG) \citep{FirstDMRG, SingleSiteDMRG}, a variational method to estimate the ground-state $|\psi_\text{gs}\rangle$ of one-dimensional Hamiltonians, although the DMRG algorithm was not originally formulated in terms of tensor networks \citep{FirstDMRG}.  
The idea is to solve the eigenvalue/eigenvector problem
\begin{equation}\label{ground_state}
H |\psi\rangle = E |\psi\rangle, 
\end{equation}
assuming that $|\psi\rangle$ is not an arbitrary quantum state, but  rather a tensor network of the form~(\ref{MPS}) with a fixed maximum bond-dimension. 
Let us denote $|\psi_\text{gs}\rangle$ the lowest energy tensor network obtained from Eq.~(\ref{ground_state}). 
This is to be contrasted with the true ground-state $|\Psi_\text{gs}\rangle$, which would be obtained if the full Hilbert space was used. 
According to the variational principle of quantum mechanics, the true ground-state energy $\mathcal{E}_\text{gs} = \langle \Psi_\text{gs} |H | \Psi_\text{gs}\rangle$ is always bounded by 
\begin{equation}\label{ground_state_energy}
E_\text{gs}:=\langle \psi_\text{gs} | H | \psi_\text{gs} \rangle \geqslant \mathcal{E}_\text{gs}.  
\end{equation}
Hence, the energy associated to $|\psi_\text{gs}\rangle$ provides an upper bound on the true ground-state energy. 

The search algorithm for the ground state is iterative. 
It proceeds by optimizing each tensor $A^{\sigma_i}_{x_{i-1}, x_{i}}$ in Eq.~(\ref{MPS}) at a time, after which it moves to the next site (details can be found in Ref. \citep{SchollwockMPS}). 
Moving one site at a time through the chain, and then backwards, is usually referred to as a sweep. 
For a fixed maximum bond dimension, multiple sweeps can be employed to ensure convergence. 
After the algorithm has converged, the bond-dimension can be increased and the process can be restarted until the desired accuracy is met. 

\section{\label{sec:HilbertSpace}Hilbert space structures for open system dynamics}

\subsection{\label{sec:vectorization}Vectorization}

We now turn to the case of open quantum systems. 
For the purpose of concreteness, we shall focus on the problem described by Eqs.~(\ref{H})-(\ref{D}). 
The generalization to other types of Hamiltonians/dissipators is straightforward. 
The master equation~(\ref{M}) is still a linear equation in $\rho$. 
The difference is that the Liouvillian $\mathcal{L}(\rho)$ is now a superoperator, as it may act on $\rho$ by means of matrix multiplications on both sides. 
This linearity can be made  manifest by introducing a vectorization operation, also called Choi-Jamiolkowski's isomorphism \citep{Choi,Jamiolkowski}, and described by 
\begin{equation}\label{vec}
\vvec\big(|i\rangle\langle j | \big) =  |j\rangle \otimes |i\rangle. 
\end{equation}
It thus converts an operator in Hilbert space, into a ket in a space whose size is the squared of the initial one. 
Matrix-wise, this corresponds to stacking the columns of a matrix, 
\begin{equation}\label{stacking_columns}
\vvec \begin{pmatrix}
a & b \\ c & d 
\end{pmatrix} = 
\begin{pmatrix} 
a \\ c \\ b \\ d
\end{pmatrix}. 
\end{equation}
Using vectorization, a general density matrix $\rho= \sum_{ij} \rho_{ij} |i\rangle\langle j |$ is converted into a ket
\begin{equation}
\vvec(\rho) = \sum\limits_{ij} \rho_{ij} |j\rangle \otimes |i \rangle. 
\end{equation}
The fact that the size of the Hilbert space is squared reflects the fact that superoperators can act on both sides of a density matrix. 
Indeed, for any 3 matrices $A$, $\rho$ and $B$, one may verify that 
\begin{equation}\label{vec_recipe}
\vvec( A\rho B) = (B\trans \otimes A) \vvec(\rho). 
\end{equation}

With this  the master equation~(\ref{M}) can be converted into a linear matrix-vector equation 
\begin{equation}\label{M_vec}
\frac{d}{dt} \vvec(\rho) = \hat{\mathcal{L}}\vvec(\rho), 
\end{equation}
where $\hat{\mathcal{L}}$ is now a matrix with entries 
\begin{equation}
\hat{\mathcal{L}} = -i (I \otimes H - H\trans \otimes I) + \hat{D}_1 + \hat{D}_N, 
\end{equation}
with $\hat{D}_i$ given by 
\begin{equation}\label{D_vec}
\hat{D}_i = \gamma_i f_i \hat{\mathcal{D}}[\sigma_i^-] + \gamma_i (1-f_i) \hat{\mathcal{D}}[\sigma_i^+], \qquad \quad i = 1,N,
\end{equation}
and with 
\begin{equation}\label{vectorized_D}
\hat{\mathcal{D}}[L] = L^* \otimes L - \frac{1}{2} \big[I \otimes (L^\dagger L )+ (L^\dagger L)\trans \otimes I\big]. 
\end{equation}
In all of the above expressions,  $I$ refers to the identity matrix with the appropriate dimension. 

The above vectorization procedure is the starting point for most numerical algorithms dealing with quantum master equations of the form~(\ref{M}). 
The  relaxation dynamics of~(\ref{M_vec}) will simply be given by 
\begin{equation}
\vvec(\rho_t) = e^{\hat{\mathcal{L}} t} \;\vvec(\rho_0), 
\end{equation}
Alternatively, one may look directly at the steady-state, Eq.~(\ref{NESS}), which now acquires the form
\begin{equation}\label{NESS_vec}
\hat{\mathcal{L}} \vvec(\rho_\text{ness}) = 0. 
\end{equation}
This equation makes explicit the fact that the NESS $\vvec(\rho_\text{ness})$ is simply the eigenvector of $\hat{\mathcal{L}}$ with eigenvalue $0$. 
Stability implies that all eigenvalues of $\hat{\mathcal{L}}$ should have non-positive real parts. 
Moreover, when the steady-state is unique, there will be only a single eigenstate with eigenvalue 0. 
One should bear in mind, notwithstanding, that the normalization of the NESS is not the standard normalization for eigenvectors. 
Instead, density operators should be normalized as $\tr(\rho) = 1$. 
This, in turn, can be viewed as the Hilbert-Schmidt inner product $\tr(A^\dagger B)$ between two operators (in this case $\rho$ and the identity matrix). 
Indeed, vectorization turns out to precisely convert Hilbert-Schmidt inner products into standard dot-products for the resulting vectors: 
\begin{equation}\label{HS_product}
\tr(A^\dagger B) = \vvec(A)^* \cdot \vvec(B).
\end{equation}
Whence, the normalization condition becomes
\begin{equation}\label{vec_normalization}
\tr(\rho) = \vvec(I) \cdot \vvec(\rho) = 1. 
\end{equation}

\subsection{Vectorization for multipartite Hilbert spaces}

\begin{figure}
	\begin{center}
	\includegraphics[trim= 0 305 0 130, clip, scale=0.3]{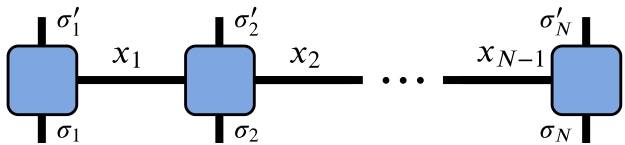}	
	\end{center}
	\caption{Upon contracting over the horizontal, indices, one recovers $\rho$.}
	\label{HilbertSpaceRho}
\end{figure}

The above recipe is not yet well suited for tensor network methods. 
The reason is that the vectorization procedure~(\ref{vec}) generally changes the tensor ordering of the Hilbert space, which can have a significant impact on the numerics. 
To see this, we consider the spin chain problem in Eqs.~(\ref{H})-(\ref{D}). 
The density matrix for this system is described by the Matrix Product Operator (MPO) 
\begin{equation}
\rho = \sum\limits_{\substack{\sigma_1,\ldots,\sigma_N \\[0.1cm] \sigma_1', \ldots, \sigma_N'}} \rho_{\sigma_1,\ldots, \sigma_N, \sigma_1', \ldots, \sigma_N'} |\sigma_1 \ldots \sigma_N \rangle\langle \sigma_1\ \ldots \sigma_N'|,
\end{equation}
where $|\sigma_1\ldots \sigma_N \rangle = |\sigma_1 \rangle \otimes \ldots \otimes |\sigma_N\rangle$ (see  Fig.~\ref{HilbertSpaceRho}).
In what follows, the tensor product symbol will be omitted for clarity. That is, we will equivalently write this as 
$|\sigma_1\ldots \sigma_N \rangle = |\sigma_1 \rangle  \ldots  |\sigma_N\rangle$.
Naive vectorization, in terms of stacking columns [Eq.~(\ref{stacking_columns})] leads to 
\begin{equation}\label{RNLN}
\vvec\big( |\sigma_1 \ldots \sigma_N \rangle\langle \sigma_1\ \ldots \sigma_N'| \big)= |\sigma_1'\rangle |\sigma_2'\rangle \ldots |\sigma_N'\rangle |\sigma_1 \rangle |\sigma_2 \rangle \ldots |\sigma_N\rangle. 
\end{equation}
Whence, we see that it rearranges the Hilbert space so as to put all right-side indices $\sigma_i'$ first, followed by all left-side indices. 
We shall refer to Eq.~(\ref{RNLN}) as the $\boldsymbol{R^N L^N}$ \textbf{ordering}.
The problem with this kind of structure, as we shall see below, is that it pushes indices pertaining to the same site, $\sigma_i$ and $\sigma_i'$, far away from each other. 

Eq.~(\ref{RNLN}) shows, in fact, that there is an  arbitrariness in how to order the Hilbert space after a vectorization. 
The order in which the indices are placed is immaterial, provided that the operators acting on $\vvec(\rho)$ are appropriately labelled to act on the correct site. 
For instance, a much more natural vectorization would be 
\begin{equation}\label{RLN}
\vvec\big( |\sigma_1 \ldots \sigma_N \rangle\langle \sigma_1\ \ldots \sigma_N'| \big)= |\sigma_1'\rangle |\sigma_1\rangle \ldots |\sigma_N'\rangle |\sigma_N \rangle,
\end{equation}
which we shall refer to as the $\boldsymbol{(RL)^N}$ \textbf{ordering}. 
This ordering preserves the ``real space'' order of the original Hilbert space, bundling together left and right indices $\sigma_i'$ and $\sigma_i$ for each site. 
Other types of orderings may also be useful, depending on the problem in question. 
Ultimately, this will depend on the kinds of operators multiplying $\vvec(\rho)$. 
As we shall see next, unitary and dissipative elements behave quite differently in this sense. 

\begin{figure}[h]
	\advance\leftskip0cm
	\subfloat[Hamiltonian interactions]{\includegraphics[trim= 0 120 10 110, clip, scale=0.17]
	{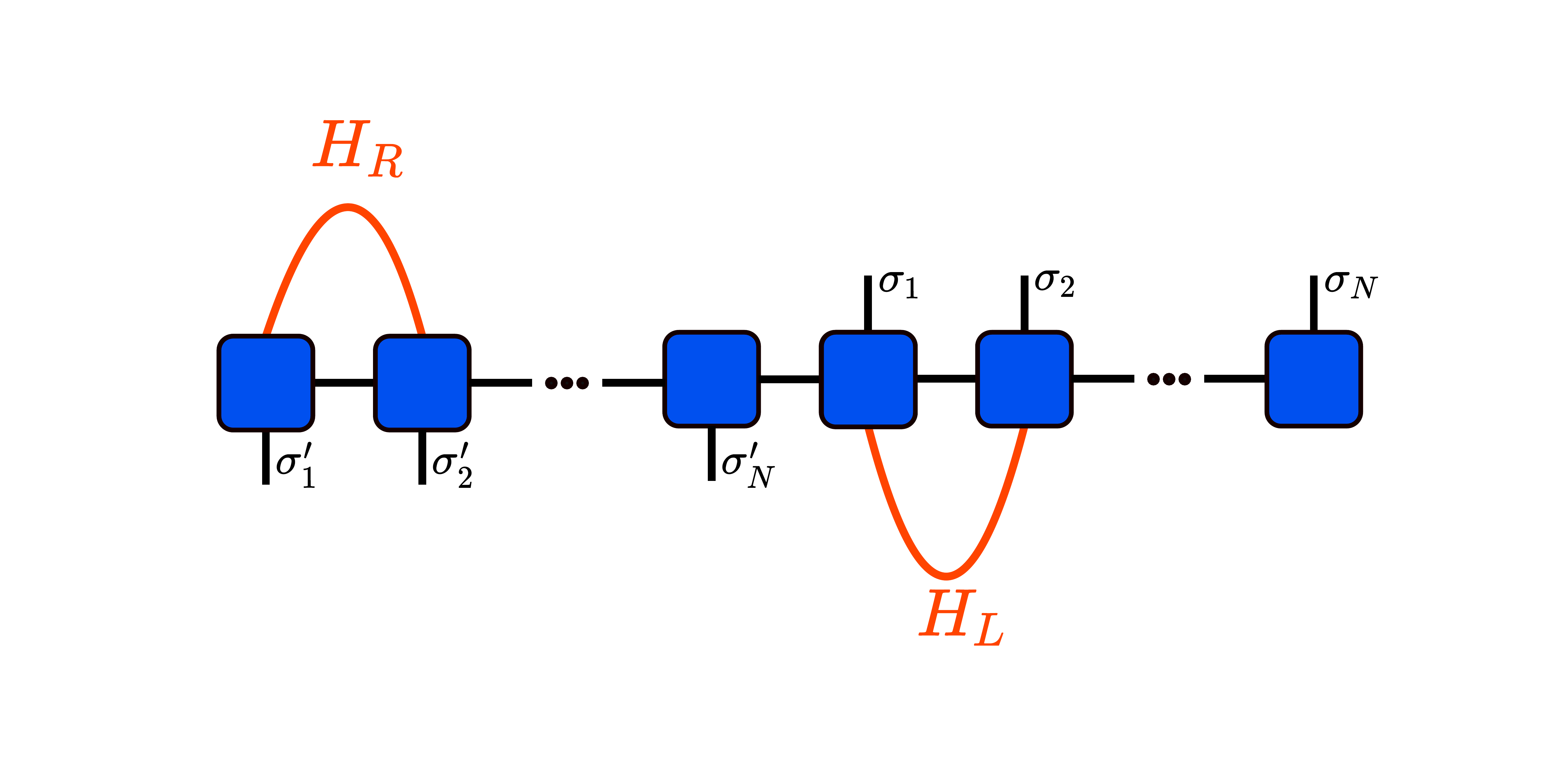}}
	\subfloat[Dissipators interactions]{
	\includegraphics[trim= 0 140 0 70, clip, scale=0.17]{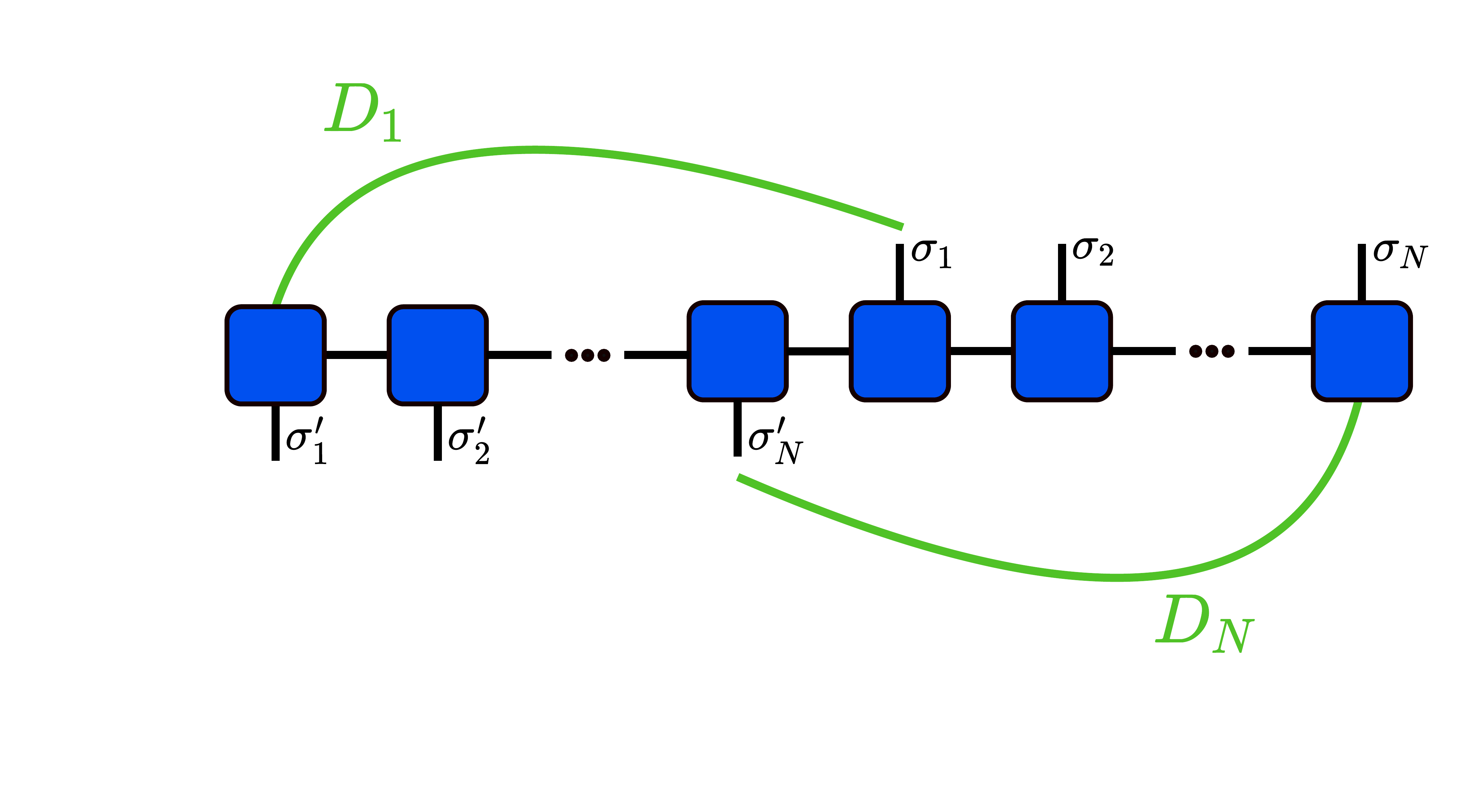}}
	\caption{Diagramatical depiction of the Hamiltonian and dissipative terms under the $\boldsymbol{R^N L^N}$ formalism.}
	\label{FigRNLN}
\end{figure}

\begin{figure}[h]
	\advance\leftskip1cm
	\subfloat[Hamiltonian interactions]{
	\includegraphics[trim= 0 210 0 110, clip, scale=0.17]{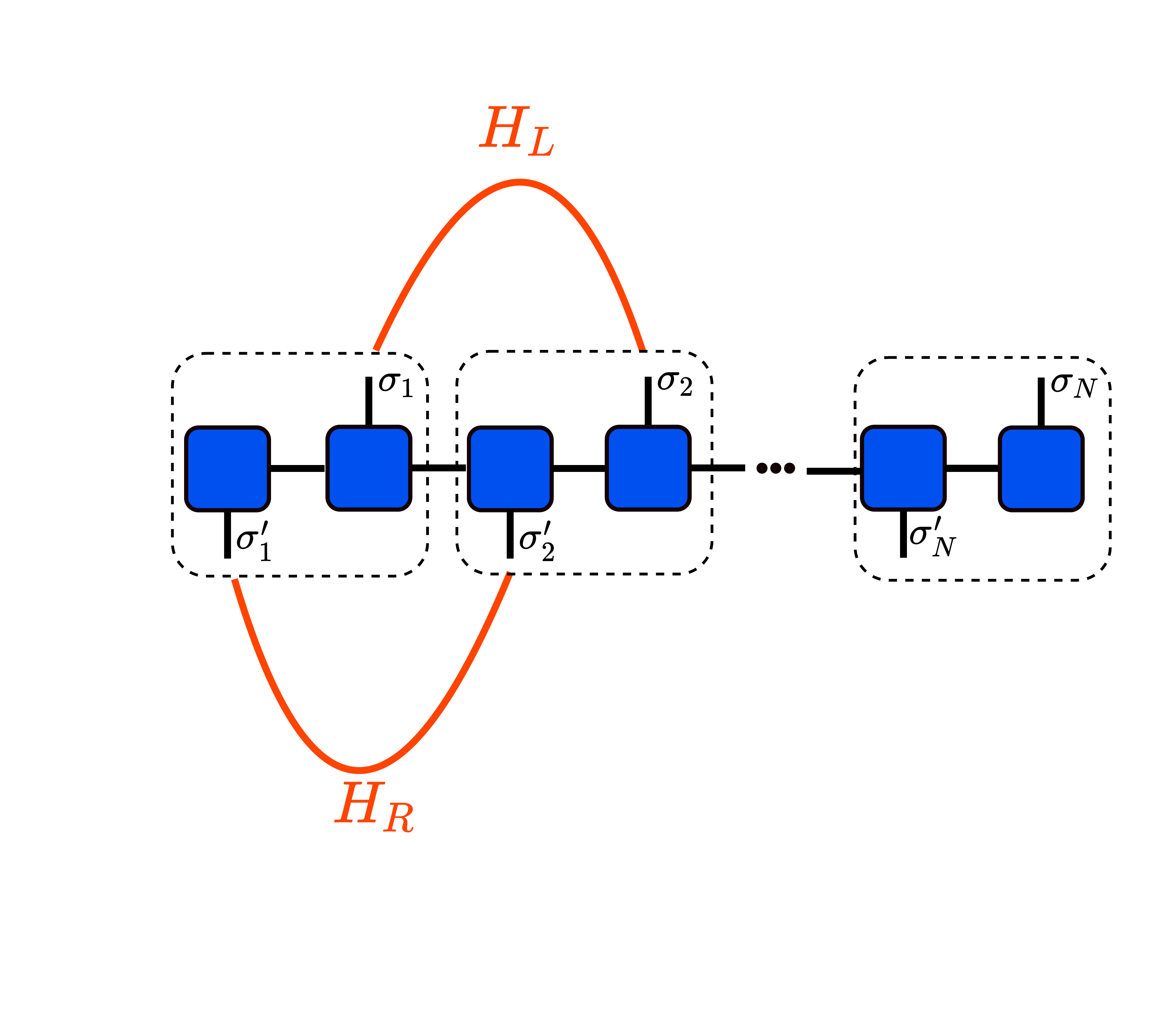}}
	\subfloat[Dissipators interactions]{
	\includegraphics[trim= 0 230 0 110, clip, scale=0.17]{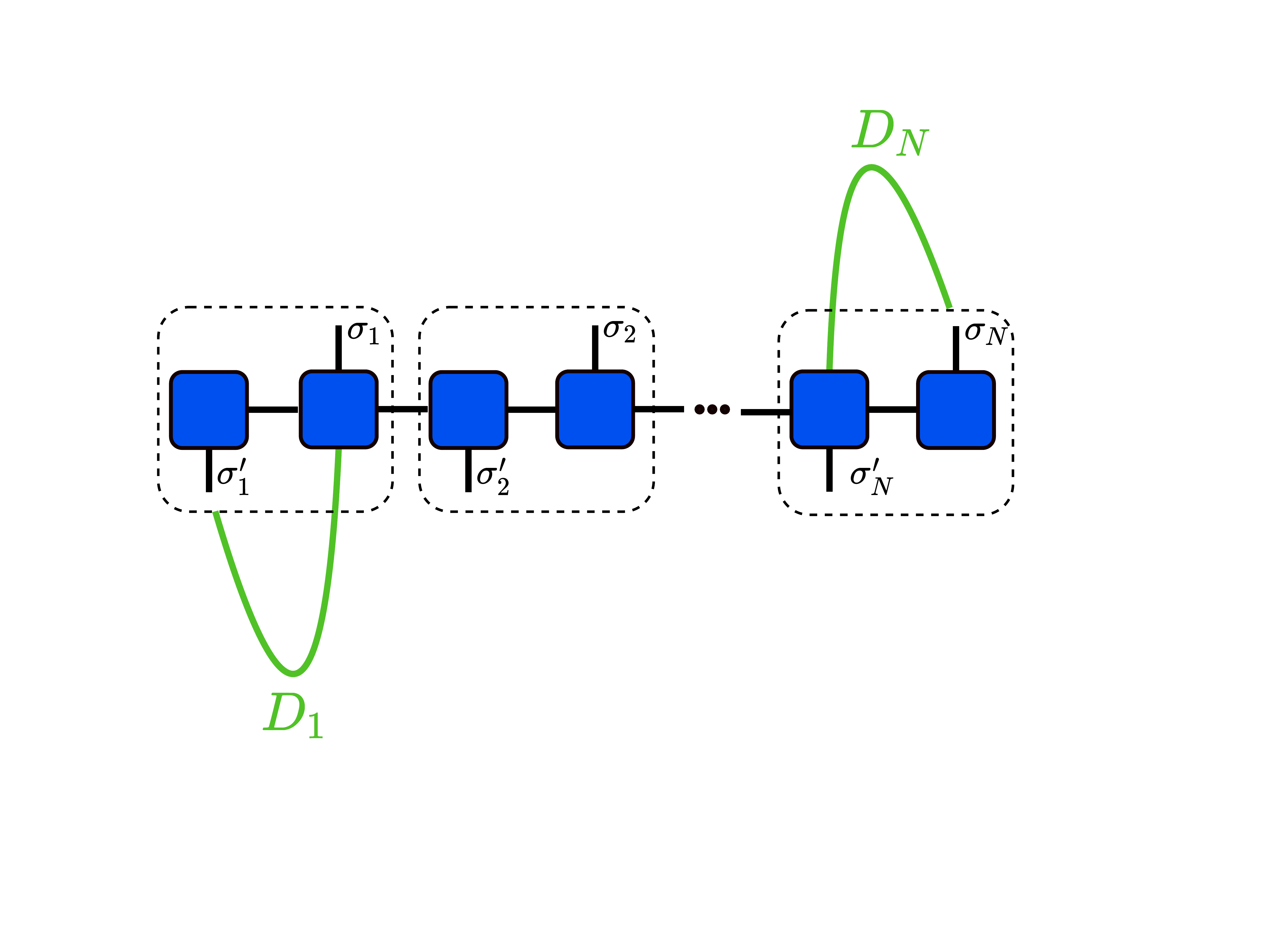}}
	\caption{Diagramatical depiction of the Hamiltonian and dissipative terms under the $\boldsymbol{(RL)^N}$ formalism. }
	\label{FigLRN}
\end{figure}

Let us begin by considering the unitary contribution. 
A typical Hamiltonian term for nearest-neighbor interactions has the form $H_1 = A_i A_{i+1}$, where $A_i$ is an operator acting on site $i$. 
This Hamiltonian will act on the master equation as $[H_1, \rho]$. 
Thus, $H_1 \rho$ will act on indices $\sigma_i$ and $\sigma_{i+1}$, whereas $\rho H_1$ will act on $\sigma_i'\sigma_{i+1}'$. 
The way this translates into the $\boldsymbol{R^N L^N}$ and $\boldsymbol{(RL)^N}$ orderings is illustrated in Figs.~\ref{FigRNLN}(a) and \ref{FigLRN}(a) respectively. 
As can be seen, in the  $\boldsymbol{R^N L^N}$ ordering the Hamiltonian retains its nearest neighbor character, with two disconnected contributions  acting on different parts of the Hilbert space. 
The $\boldsymbol{(RL)^N}$ ordering, on the other hand, leads to a second nearest neighbor interaction. 

We now move on to the dissipative contributions. 
The special part is the first term in Eq.~(\ref{vectorized_D}). A dissipator such as $\mathcal{D}[\sigma_1^-]$ for instance, has a contribution of the form $\sigma_1^- \rho \sigma_1^+$. 
This will act on indices $\sigma_1$ and $\sigma_1'$. 
The corresponding tensor structure for the two orderings will then be as shown in Figs.~\ref{FigRNLN}(b) and \ref{FigLRN}(b).
As we now see, the $\boldsymbol{R^N L^N}$ ordering leads to a {\it highly} non-local Hilbert space structure, whereas in $\boldsymbol{(RL)^N}$ the structure is nearest-neighbor. 

This analysis clearly shows why the $\boldsymbol{(RL)^N}$ ordering (Fig.~\ref{FigLRN}) will in general fare better in a numerical calculation: even though the Hamiltonian is now a second nearest-neighbor interaction, the dissipator is only nearest-neighbor. 
This will be more advantageous than $\boldsymbol{R^N L^N}$ which has long-range interacting terms.

\section{\label{sec:implementation}Implementation of the oDMRG algorithm}

\subsection{The $\mathcal{L}^\dagger \mathcal{L}$ method}

Having established the Hilbert space structure, we are now in a position to implement the oDMRG algorithm introduced in \cite{LdLTechnique}. 
The starting point is the steady-state equation~(\ref{NESS_vec}), which shows that the NESS is the eigenstate of $\hat{\mathcal{L}}$ with eigenvalue $0$. 
The problem with this equation is that $\hat{\mathcal{L}}$ is a non-Hermitian operator. 
To circumvent this, we consider instead the eigenvalue/eigenvector of $\hat{\mathcal{M}} = \hat{\mathcal{L}}^\dagger \hat{\mathcal{L}}$~\cite{LdLTechnique}; that is, instead of~(\ref{NESS_vec}) we solve
\begin{equation}\label{M_LdL}
	\hat{\mathcal{M}}  \vvec(\rho_\text{ness}) = \hat{\mathcal{L}}^\dagger \hat{\mathcal{L}}\; \vvec(\rho_\text{ness})  = 0. 
\end{equation}
The operator $\hat{\mathcal{M}}$ has the same steady-state as $\hat{\mathcal{L}}$, but is  Hermitian. 
Moreover, $\hat{\mathcal{M}}$ is by construction positive semi-definite, with exactly one zero eigenvalue (when the steady-state is unique) and all other eigenvalues strictly larger than zero. 

For these reasons, Eq.~(\ref{M_LdL}) has now the exact same structure as the closed system eigenvalue problem Eq.~(\ref{ground_state}): we need essentially to look for the ground-state of the effective Hamiltonian given by $\hat{\mathcal{M}}$, and the search for this groundstate is therefore entirely amenable to the closed DMRG algorithm. 
Eq.~(\ref{M_LdL}) also offers the additional advantage that the ground-state energy is known exactly, $E_\text{gs} = 0$. 
Hence, monitoring how the energy changes during the DMRG sweeps can be used as a way to probe the convergence of the algorithm (for an implementation of a DMRG-like code for the non-Hermitian superoperator $\hat{\mathcal{L}}$ see \citep{MascarenhasSavona2015}). 

For concreteness, we shall henceforth focus on the model in Eqs.~(\ref{H})-(\ref{D}) and choose the initial parameters such that $J_i = 1$, $\Delta_i = \Delta$ and $h_i = h$. 
The input parameters are then only the chain size $N$, together with $\gamma$, $f_1$, $f_N$, $h$ and $\Delta$.

One setback of the algorithm~(\ref{M_LdL}) is that even if the $\boldsymbol{(RL)^N}$ ordering is used for $\hat{\mathcal{L}}$, the tensor structure of $\hat{\mathcal{M}} = \hat{\mathcal{L}}^\dagger \hat{\mathcal{L}}$ will now be highly non-local. 
This should be expected because a steady state of local Hamiltonian and dissipative terms, in general does not need to follow an area-law, unlike the ground state of local Hamiltonians. 
However the non-locality of the terms may lead to an ``entanglement-barrier'' problem for the convergence of the algorithm. 
The reason is that, during convergence, the algorithm will pass through multiple, highly entangled, eigenstates of the operator $\hat{\mathcal{M}}$. As a side comment, note that these excited states are typically different from the eigenstates of $\hat{\mathcal{L}}$ because non-Hermitian operators have different left and right eigenvectors (the NESS is an eigenstate that both operators share). 
Another issue is that $\hat{\mathcal{M}}$ has usually a smaller gap, between the steady state and the first excited state, as compared to $\hat{\mathcal{L}}$, making the problem numerically harder to converge. 
These problems were investigated recently in Ref.~\cite{Gangat2017}, where the authors proposed additional approximations for making $\hat{\mathcal{M}}$ more local. 
In \cite{Gangat2017}, however, the authors did not deal with boundary-driven transport problems. 
In our implementation we will consider the full $\hat{\mathcal{M}}$ operator, and as we show, for the boundary-driven problems we studied, good convergence rates were  obtained without the need for these additional methods.

\subsection{Positivity of the variational density matrix}

Another side effect of using a closed-system algorithm for open quantum systems concerns the positivity of the numerically obtained density matrix $\vvec(\rho)$. 
As discussed in Sec.~\ref{sec:vectorization}, the vectorized density matrix is not normalized as a standard vector, but rather as in Eq.~(\ref{vec_normalization}).  
This is at odds with the closed DMRG algorithm, which uses standard normalization. 
This, of course, can be readily fixed by appropriately renormalizing the output state. 
A more serious issue, however, concerns the positivity of the resulting density matrix: physical density matrices must be positive semi-definite. 
A physical tensor network variational state for $\vvec(\rho)$ must therefore be one for which the corresponding ``unvec'd'' state is positive semi-definite. 
The set of tensor network states through which the system passes  during the algorithm, however, is not restricted to this, but may very well contain also non-positive states. 
The set is also not convex, so that even if we were to start with a physical state, there is no guarantee that the algorithm remains in one. 
As a consequence, it is possible that the algorithm~(\ref{M_LdL}) converges to states which have low energies but are otherwise unphysical. 
This can be witnessed, for instance, by imaginary contributions to the expected values of observables. 

We have found that this problem can be dramatically minimized by adopting the following procedure. 
First, we use as the starting guess for the tensor network state, a maximally mixed state, $\vvec(I)$, an object which below we refer to as \verb+Ivec+.
Second, we start the process with very small bond dimensions, usually 2. 
Such a small bond dimension allows for extremely fast computations, so that we allow for a large number of sweeps to ensure convergence. 
For such a small bond-dimension, the system is found to  naturally converge to a physical state. 
We call this first phase the {\it warm-up}. 
Finally, and most importantly, we then proceed to increase the bond dimension in very small steps, usually in steps of 1 or 2 (allowing, of course, multiple sweeps for each bond dimension to ensure convergence).
The reason why this works is because if the bond-dimension is too large, the algorithm will generally converge towards unphysical states. 
But by incrementing the bond dimension in small steps, one minimizes these disturbances, pushing the system towards the manifold of positive semi-definite states.
We have no formal proof that this approach necessarily has to work. But in all scenarios we have tested, it was found to  dramatically improve the results.

\subsection{\label{Comparison}Numerical comparison between the $\boldsymbol{(RL)^N}$ and $\boldsymbol{R^NL^N}$ orderings}

\begin{figure}[h]
	\centering
	\includegraphics[trim= 0 0 0 0, clip, scale=0.3]{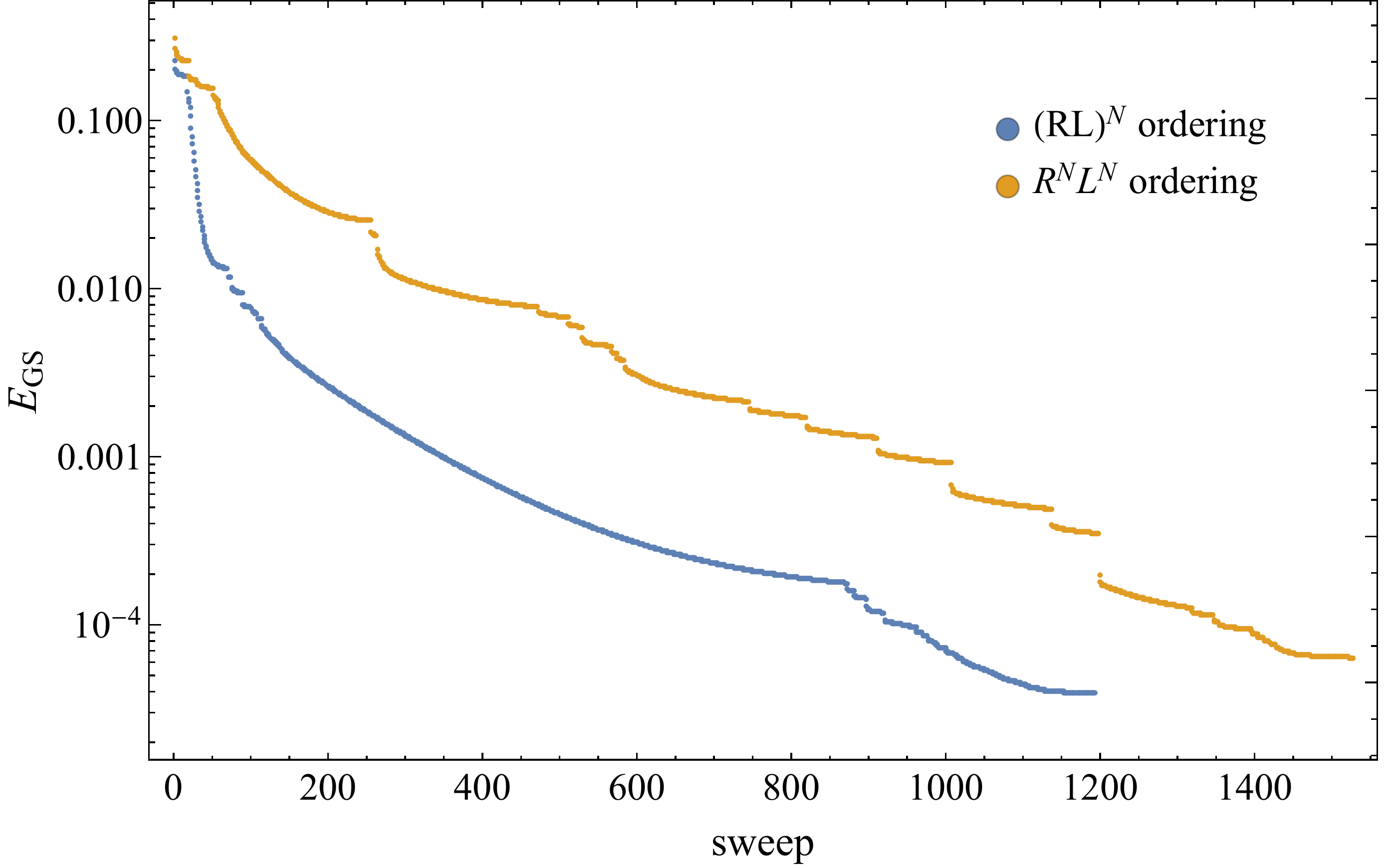}
	\caption{Energy as a function of the number of sweeps, for the $\boldsymbol{(RL)^N}$ and  $\boldsymbol{R^NL^N}$ orderings, with a fixed maximum bond dimension of 100. Values used: $N = 20, \gamma = 1, \Delta = 0.5, f_1 = 0.8, f_N = 0.2, h = 0$. }
	\label{RLN-RNLN}
\end{figure}

We implement the above steps using the iTensor library~\cite{ITensor}. 
Initially, for the sake of comparison, we have implemented both the $\boldsymbol{(RL)^N}$ and  $\boldsymbol{R^NL^N}$ orderings, and we provide in Fig.~\ref{RLN-RNLN} a convergence test for both.
This is done by monitoring the lowest eigenvalue of $\mathcal{M}=\mathcal{L}^\dagger \mathcal{L}$ which, as already discussed, is called  ``energy'', in reference to Eq.~\eqref{ground_state_energy}. 
Recall that in our case the energy of the true steady-state is known to be identically zero. 
Thus, its magnitude  serves as a quantifier of the convergence of the algorithm. 
Indeed, as the plot indicates, the $\boldsymbol{(RL)^N}$ ordering is consistently more reliable, having a smoother convergence curve after each  sweep, and requiring less sweeps to achieve better numerical results.
In light  of this, the discussion below will be centered on the $\boldsymbol{(RL)^N}$ ordering.
%

%

\subsection{\label{initCode}Initialization}

We now discuss the details of the implementation, focusing on the code available at \cite{oDMRG}.
Further details on the functions presented in this section can be found in \ref{sec:app}. 
The tensor class is called by
\begin{lstlisting}[frame=single,language=C++]
     auto sites = LRN(N);
     MPS rho = MPS(sites);
\end{lstlisting}
The resulting object \verb+sites+, which is the output of the function \verb+LRN+, contains all definitions of the Hilbert space structure, together with how the Pauli operators act on different indices for left- and right-multiplication.

All observables can then be constructed from the \verb+sites+ object. 
Here we focus on the currents~(\ref{current}) and the local magnetization $\sigma_z^i$. 
iTensor allows for a simple symbolic input for building operators, which we have adapted to include left and right multiplication. 
The resulting code is then implemented as
\begin{lstlisting}[frame=single,language=C++]
    for(int i=1; i<N; i++){
        auto aobs  = AutoMPO(sites);
        aobs +=  4.0,"SxL",i,"SyL",i+1;
        aobs += -4.0,"SyL",i,"SxL",i+1;
  	    obsCurrVec.push_back(MPO(aobs));
    }   
    for(int i=1; i<=N; i++){
        auto aobs  = AutoMPO(sites);
        aobs += 2.0,"SzL",i;
        obsMagVec.push_back(MPO(aobs));
    } 
\end{lstlisting}	
An object such as \verb+``SxL''+, for instance, stands for the Pauli matrix acting on the left. 
Similarly,  \texttt{``SxL'',i,``SyL'',i+1} stand for the operator $\sigma_x^i \sigma_y^{i+1}$ acting on the left. 
The factors of $4$ and $2$ are simply because iTensor naturally loads spin operators $S_x^i = \sigma_x^i/2$, etc. 
The above routine constructs a list of MPOs, each representing the current in a given bond or the magnetization in a given site. 

Next we construct the tensor network for the density matrix $\vvec(\rho)$, which is the object that will be optimized in the algorithm. 
We also initialize it to the maximally mixed state $\vvec(I)$ (normalization is not required and is done only when we compute the expectation values of observables). 
The code reads
\begin{lstlisting}[frame=single]
     MPS rho;
     MakeIVEC(rho, N);
\end{lstlisting}     
The function \verb+MakeIVEC+ constructs an MPO of the form $\text{vec}(I)$.
Lastly, we construct the matrix $\hat{\mathcal{M}} = \hat{\mathcal{L}}^\dagger \hat{\mathcal{L}}$:
\begin{lstlisting}[frame=single]
    MPO LdL = LdLXXZConstruct(sites, Delta, f1, fL, gamma, h);
\end{lstlisting}
This functions uses the same type of constructs used in iTensor to build Hamiltonians, but again taking care of proper left and right multiplications. 
It also uses a symbolic structure  to construct the object in a way that is independent of the bond dimension being used. 
As a consequence, the resulting object \verb+LdL+ has no significant memory cost, irrespective of the size $N$. 
We also mention that while the above function focuses on a homogenous chain (i.e. homogeneous $\Delta$ and $h$), it is trivial to extend it to the inhomogeneous case.

\subsection{Warm-up}
\label{WarmUpSec}

\begin{figure}[h]
	\advance\leftskip-0.5cm
	\subfloat[Ground-state energy guess after each DMRG sweep]{
	\includegraphics[width = 3.3in]{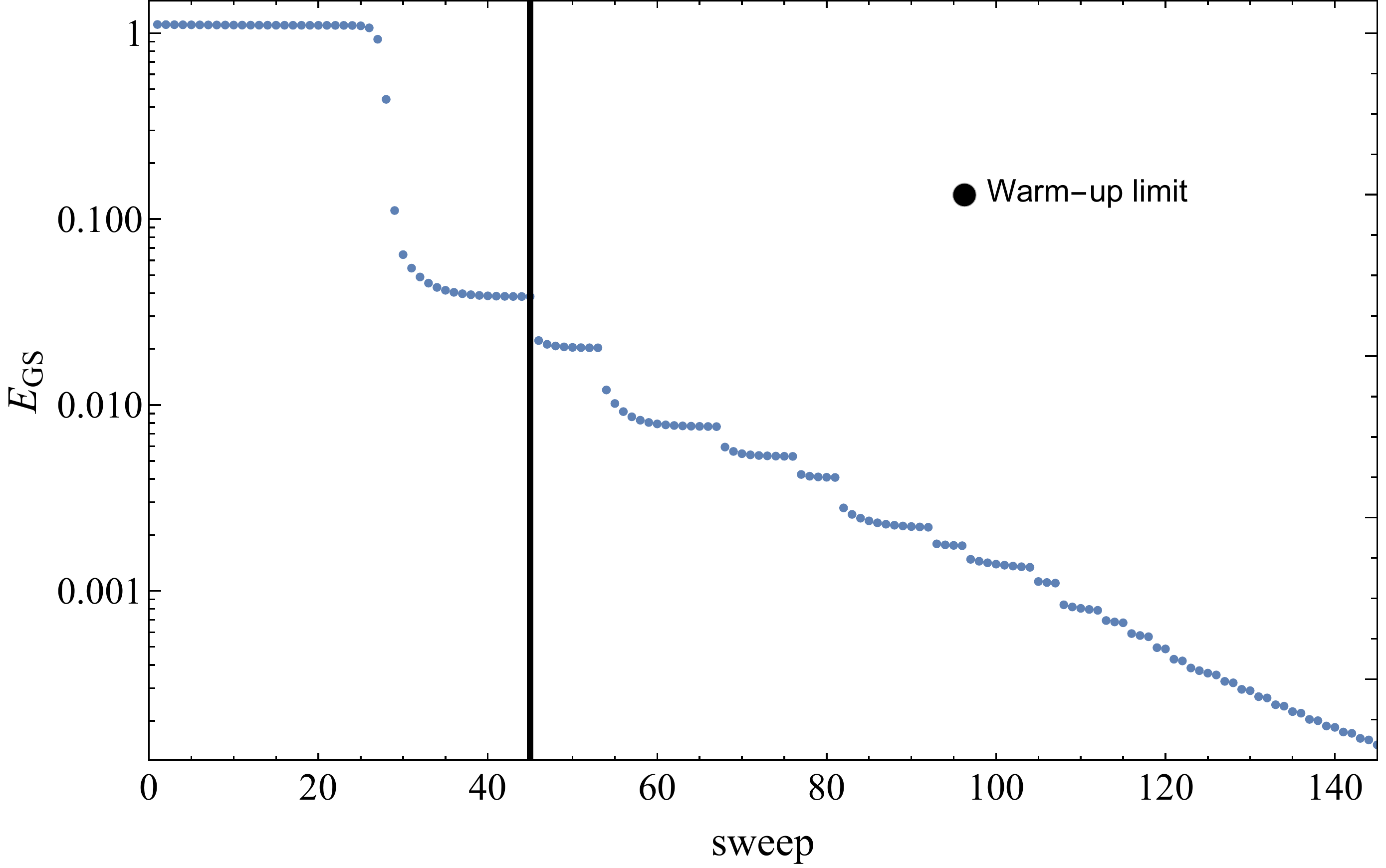}
	}
	\subfloat[Wall-time for each sweep]{
	\includegraphics[width = 3.3in]{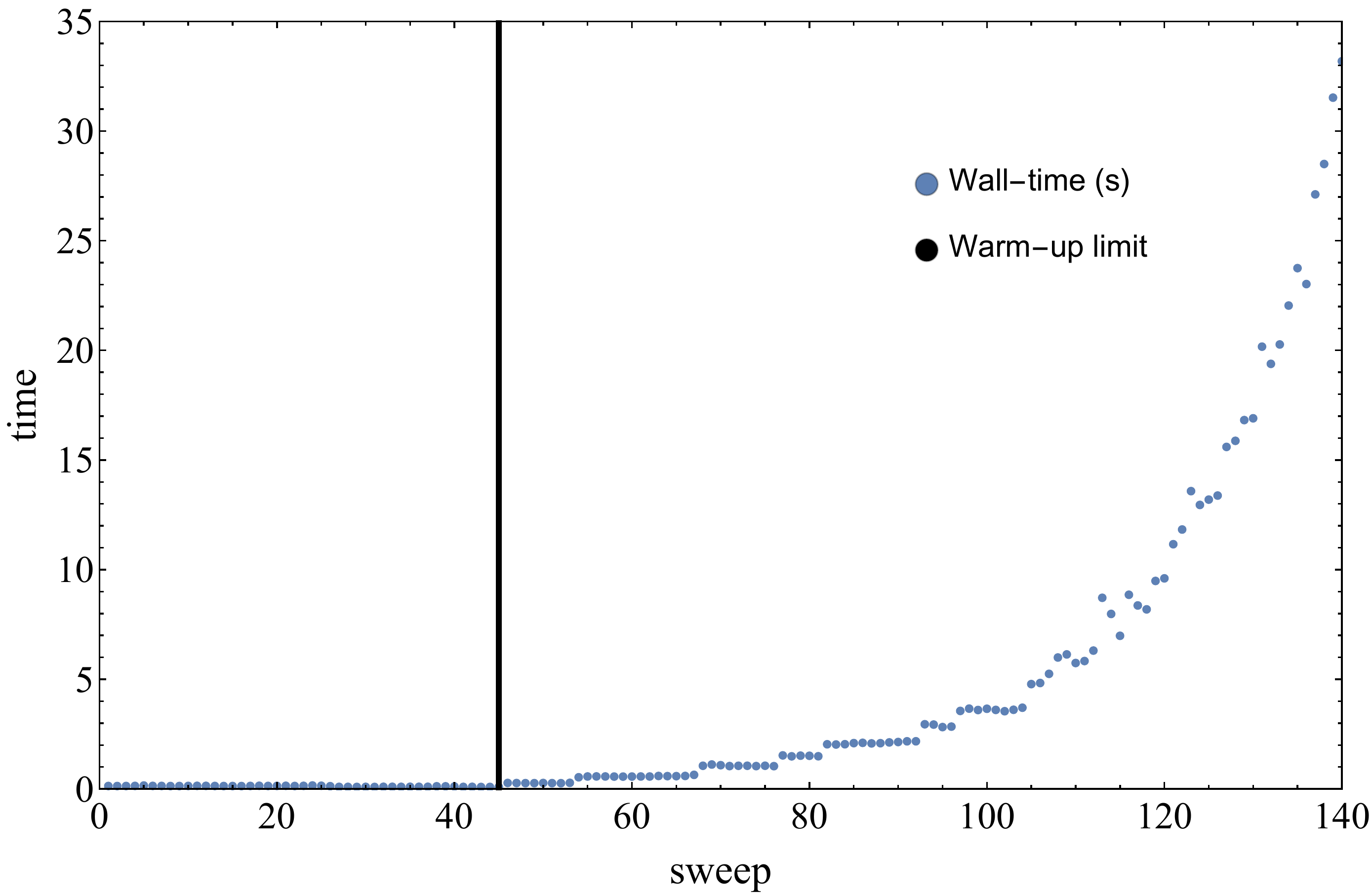}
	}\\
	\caption[Plots of both ground state energy and time taken by sweep per sweep.]{Plots of both ground state energy and time taken by sweep as a function of the number of sweeps. The black horizontal line represents the sweep where the warm-up ends and the actual simulation begins.
	Values used: $N = 10, \gamma = 1, f_1 = 1, f_L = 0, h=0, \Delta = 1$.}
	\label{WarmUpTime}
\end{figure}

As discussed above, we perform a warm-up routine to improve the convergence to a physical tensor network. 
It performs multiple DMRG-sweeps with the lowest bond-dimension, to sharpen the initial parts of the simulation. The function receives the state \verb+rho+, the MPO object \verb+LdL+, the error threshold to stop the function, and an additional tag (which can be either "true" of "false") to manage the output of the function to the console.
\begin{lstlisting}[frame=single]
    WarmUp(rho, LdL, 0.001, {"Quiet", true});
\end{lstlisting}
This function is a minor adaptation of iTensors built-in DMRG routine. 
The improvements brought about by the warm-up are significant, as shown in Fig.~\ref{WarmUpTime}.
 The black horizontal line represents the sweep where the warm-up ends and the actual simulation begins (to be discussed in what follows).
 As can be seen, the reduction in energy during the warm-up is significant, even though the simulation time [Fig.~\ref{WarmUpTime}(b)] is negligibly small. 

\subsection{DMRG sweeps and final calculations}

After the initial warm-up routine, the DMRG procedure is then applied for increasing values of the bond-dimension parameter. 
The individual DMRG runs are called as
\begin{lstlisting}[frame=single]
energyFin = dmrg(rho,LdL,sweeps});  
\end{lstlisting}
which is just a call to the built-in DMRG function from iTensor.
This can then be placed inside a loop, which compares the energy  with the previous value; if the two  fall within 10\% of one another, the bond-dimension value is  increased by a fixed amount. Both of these parameters, the threshold upon which one increases the bond dimension, and the amount of the bond dimension increase, can be easily altered by the user at the initial lines of the main routine. The algorithm can be run indefinitely, or the user may choose a stopping point, for instance, the maximum bond dimension, amount of sweeps. etc.

During each sweep, we calculate the spin current Eq.({\ref{current}}) throughout the chain, as well as the magnetization for each site. Both of these are done in a similar manner, contracting the previously loaded MPO of the  relevant site with  the tensor network object $\rho$. We take advantage of iTensor's  optimized tensor network procedures. The value is then printed out. For example, the magnetization is computed with a loop going up to the size of the chain, for each site calculating
\begin{equation}\label{M_LdL}
\dfrac{\langle I_{vec} | M_{j} | \rho \rangle}{\langle I_{vec} | \rho \rangle},
\end{equation}
where $M_j$ is the magnetization MPO described in Sec.~\ref{initCode}. 
This is  done through the following excerpt:
\begin{lstlisting}[frame=single]
for(int j=1;j<=N;j++){
	  auto ev = overlapC(Ivec,obsMagVec[j-1],rho)/overlapC(Ivec,rho);
}  
\end{lstlisting}
A similar procedure is done  for the spin current.
\begin{lstlisting}[frame=single]
for(int j=1; j<N; j++){
    auto current = overlapC(Ivec,obsCurrVec[j-1],rho)/overlapC(Ivec,rho);
}
\end{lstlisting}
In both cases we use the previously calculated vector of MPOs from Sec.\ref{initCode}.
Of course, to optimize the code, one may also only compute the observables at the end of the process. 
Here we compute them at each step in order to monitor their convergence. 

Finally, at the end of each sweep, we check if the energy has stabilized, and, if so, we increase the bond dimension value in order to advance the accuracy of the routine.
\begin{lstlisting}[frame=single]
if ((energyIni-energyFin)/energyIni < sweepBDChangeThresholdValue){
	bd += BDinc;
	sweeps.maxm() = bd;
}
\end{lstlisting}
Additional conditions can be easily implemented taking into account the simulation parameters, to finely tune the convergence of a specific set of parameters within a specific system.

\section{\label{sec:results}Results}

\subsection{Benchmarking convergence}

\begin{figure}
\centering
		\includegraphics[trim= 0 0 0 0, clip, scale=0.3]{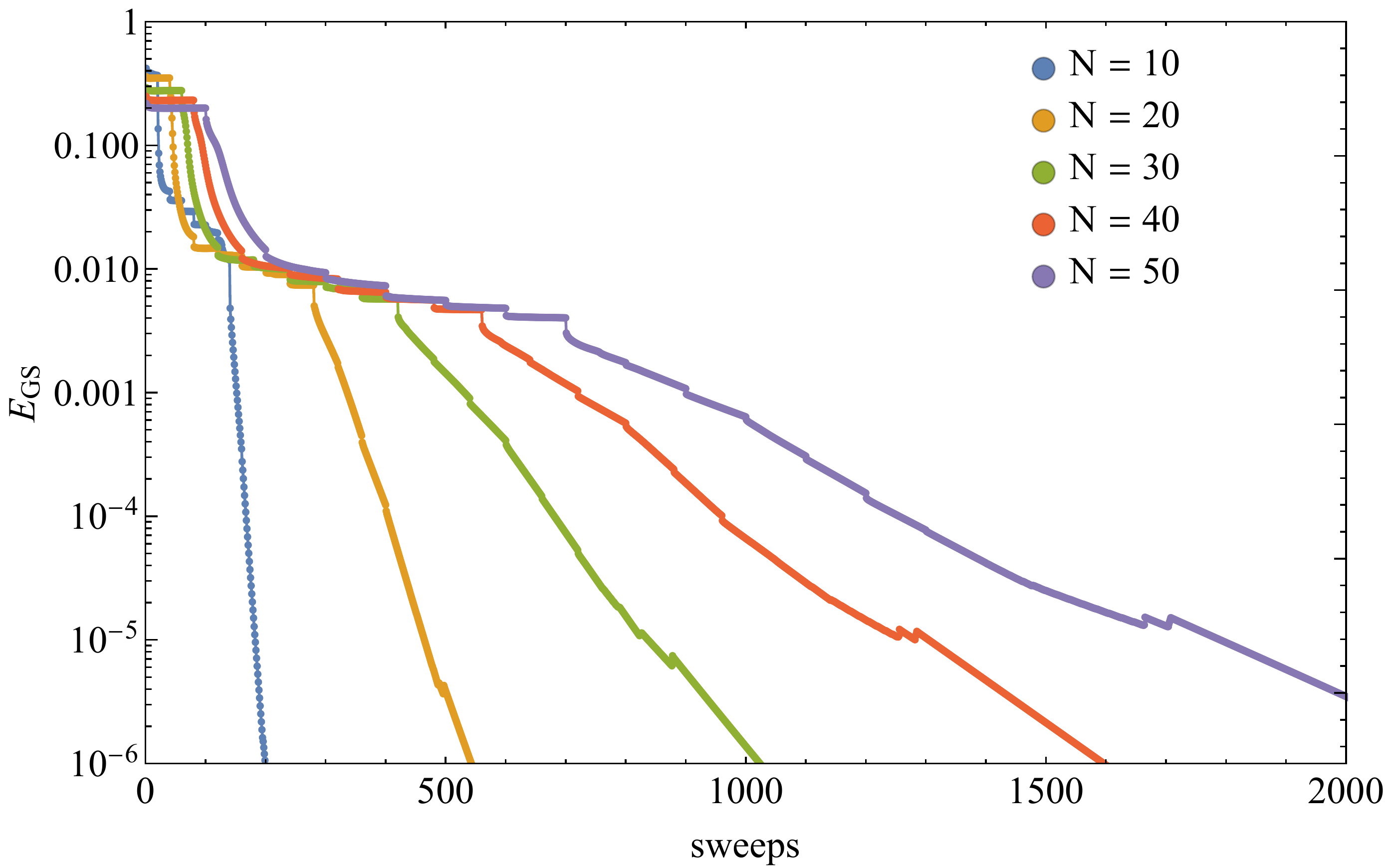}
	\caption{Energy convergence for different chain lengths, after each sweep, for different system sizes $N$. Values used: $\gamma = 1, \Delta = 0.5, f_1 = 1, f_N = 0, h = 0$. }
	\label{GSConvergence-N10}
\end{figure}

\begin{figure}
\centering
	\includegraphics[trim= 0 0 0 0, clip, scale=0.3]{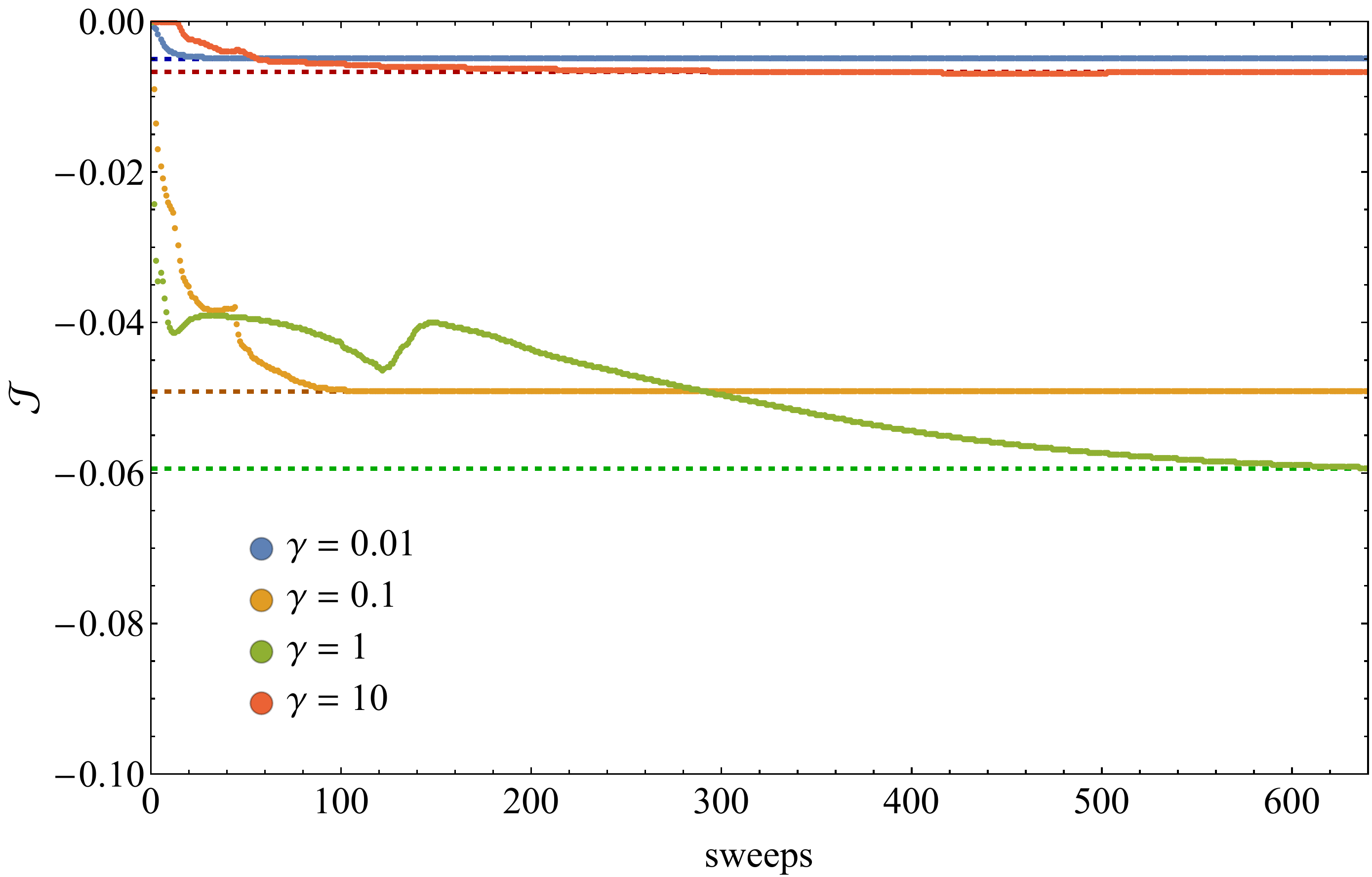}
	\caption{Spin current convergence versus sweep number for different coupling values $\gamma$. Values used: $N = 25, \Delta = 1, f_1 = 1, f_N = 0, h = 0$. Dashed lines correspond to the analytical solutions in~\citep{Landi2015}.}
	\label{FluxConvergence}
\end{figure}

Initially, we look at the convergence of the algorithm for different chain sizes.
Illustrative results are shown in Fig.~\ref{GSConvergence-N10} for sizes up to $N = 50$. 
As can be seen, for small sizes the convergence is extremely fast. 
Increasing the size of the chain makes it so that more sweeps are necessary, but since the bond dimension of each sweep is increased in a slow, controlled manner, the convergence is possible even for larger sizes.
For the particular case of $\Delta = 1$, $f_1 = 1$ and $f_N = 0$, the problem actually has an analytical solution in the form of a matrix product ansatz~\citep{Landi2015, Prosen2011b}.
By looking at the average current after each sweep, we can therefore  benchmark the algorithm to assure the convergence of the current.
This is illustrated in Fig.~\ref{FluxConvergence}. 
As can be seen, the convergence is generally slower for intermediate values of $\gamma$. 
Taking the analytical values available as references, we can see that the algorithm is working as intended. Additionally, all these simulations  were made in the span of a couple days, with an average desktop: no broad computational power was required.

\subsection{Benchmarking the steady-state in comparison with analytical solutions}

\begin{figure}
\centering
	\subfloat[]{\includegraphics[width = 3in]{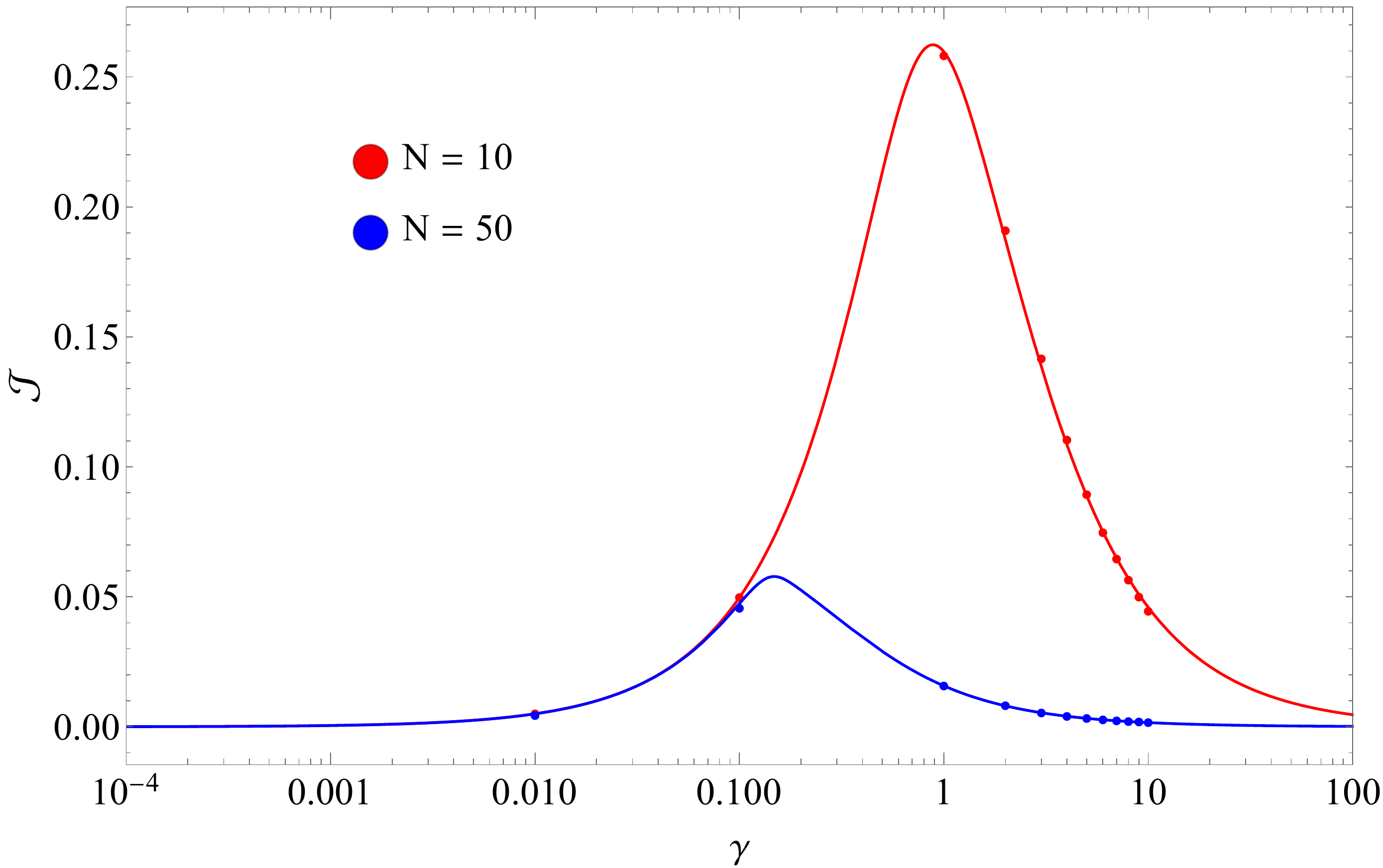}} 	\qquad
	\subfloat[]{\includegraphics[width = 3in]{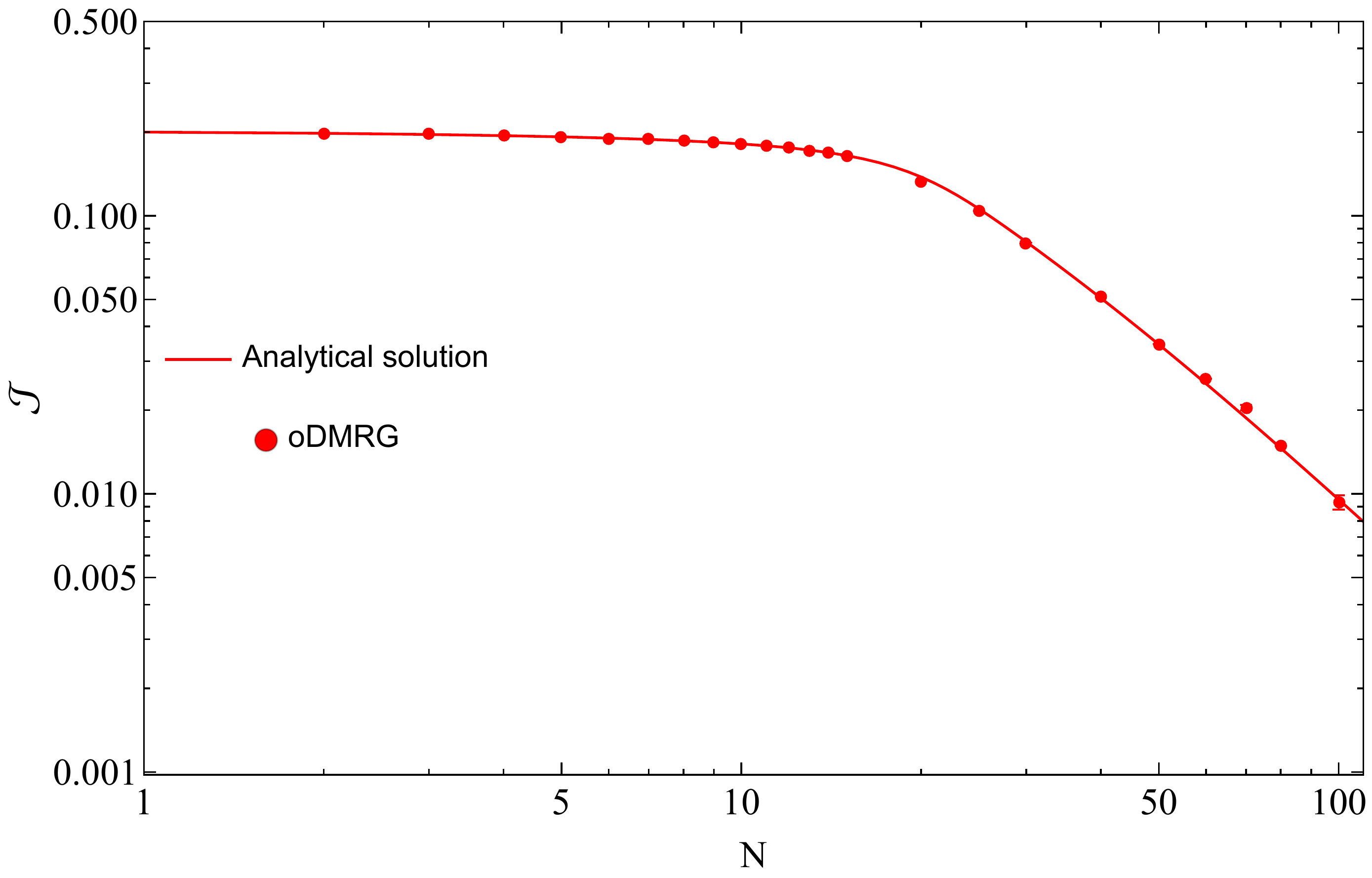}} 	
	\caption{(a) $\mathcal{J}$~vs.~$\gamma$ for two values of $L$.  The solid line corresponds to the analytical solution from~\citep{Landi2015} and the dots correspond to the oDMRG simulations.
	(b) $\mathcal{J}$~vs.~$N$ for $\gamma = 0.5$.
	One can clearly see the change from ballistic to subdiffusive transport.
	Other parameters were $\Delta = 1, f_1 = 1, f_N = 0, h = 0$.
	}
	\label{SpinFluxAnalytical}
\end{figure}

\begin{figure}
	\subfloat[]{\includegraphics[width = 3in]{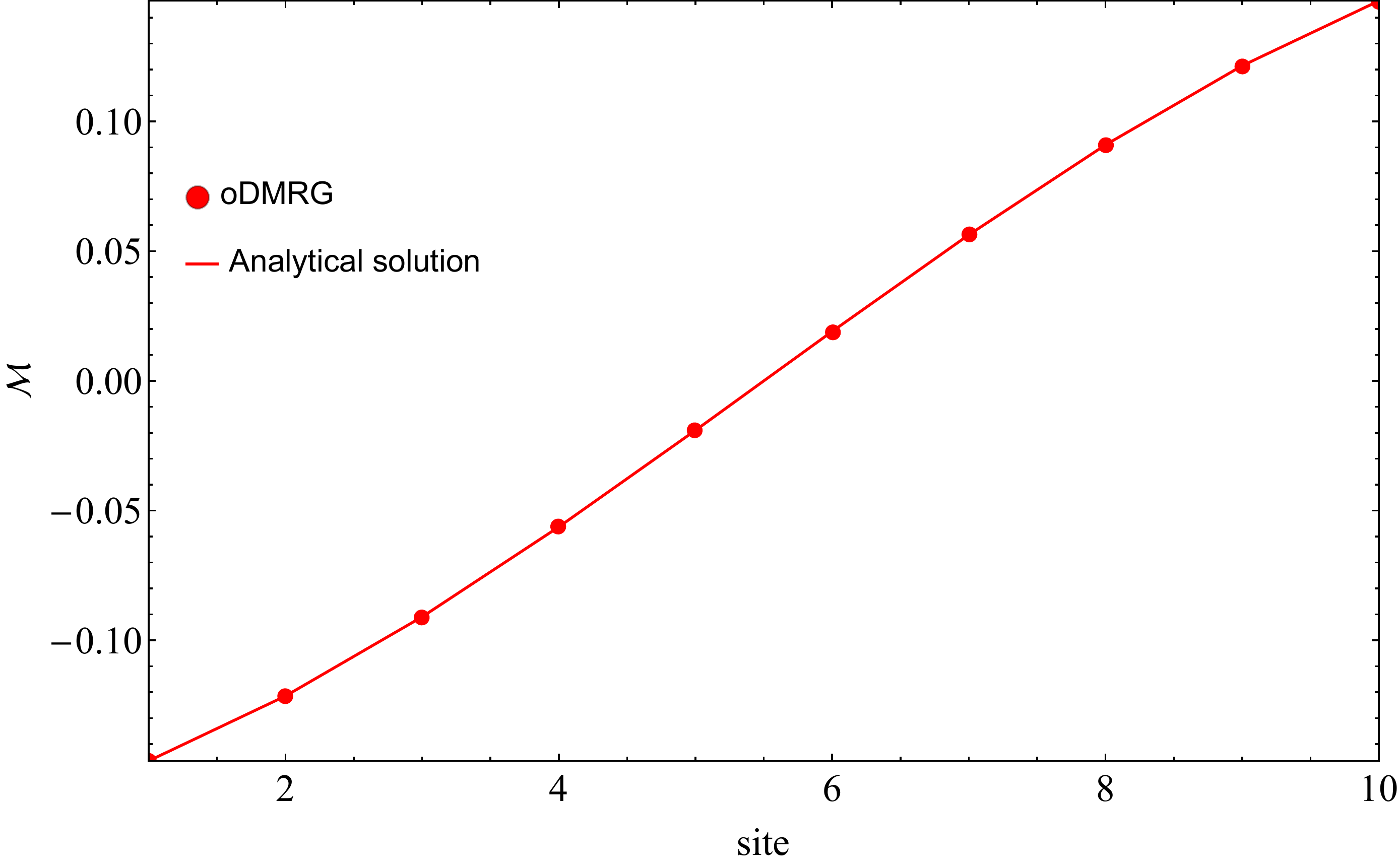}} \qquad 
	\subfloat[]{\includegraphics[width = 3in]{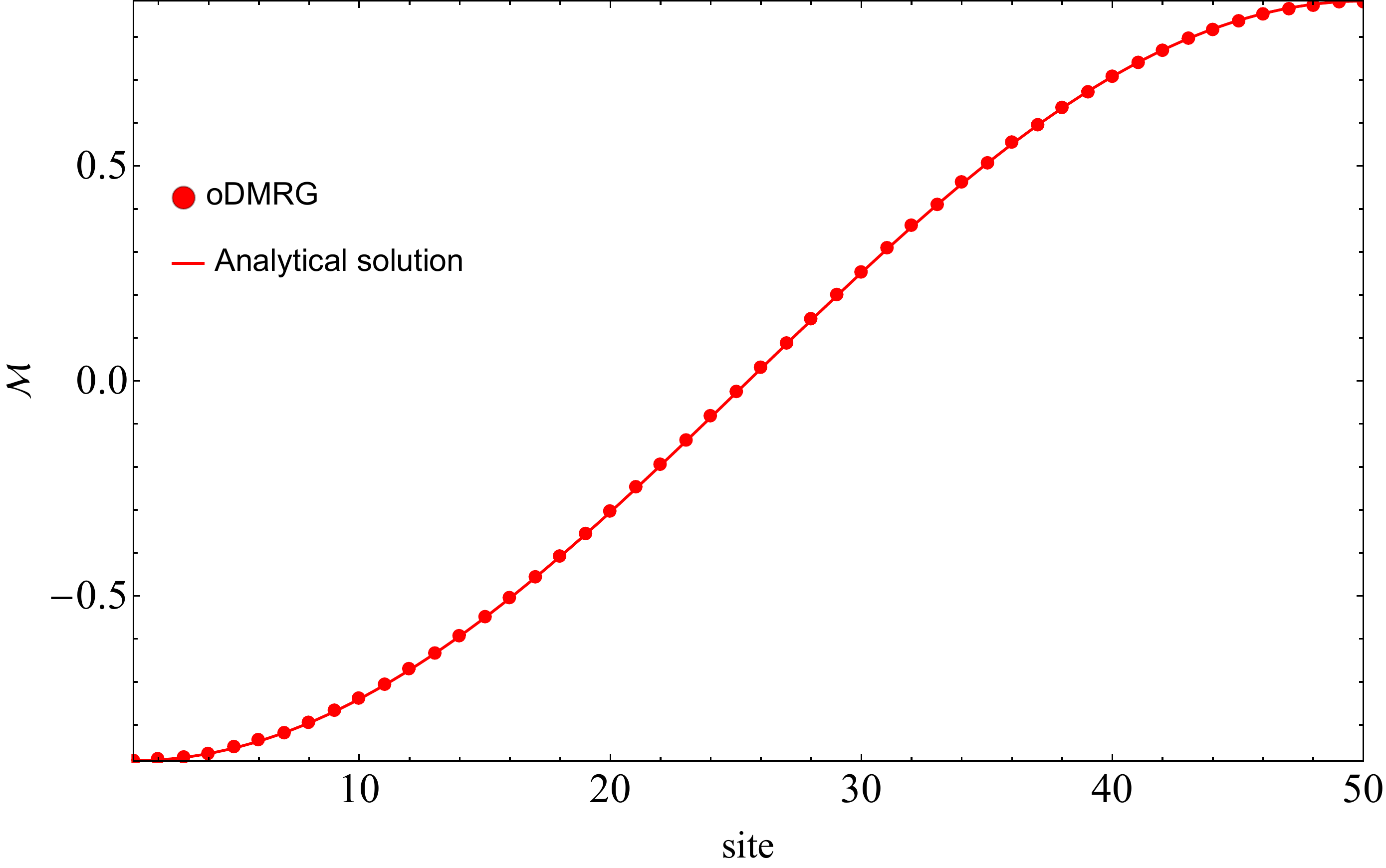}}	
	\caption{Magnetization profile $\mathcal{M} = \langle \sigma_z^i \rangle$ as a function of the sites for $\gamma = 0.5$, for (a) $N = 10$  and (b) $N = 50$. 
	The red curves are the exact solutions from~\citep{Landi2015}. 
	Other parameters were $\Delta = 1, f_1 = 1, f_N = 0, h = 0$.}
	\label{MagneticProfileAnalytic}
\end{figure}

After assuring that the code is working, we can further our analysis by studying the current for different coupling values $\gamma$, for two chain sizes.
The steady-state current as a function of $\gamma$ is shown in Fig.~\ref{SpinFluxAnalytical}(a), where it is compared with the analytical solution (solid lines). 
As can be seen, the agreement is extremely good. 
Similarly, in Fig.~\ref{SpinFluxAnalytical}(b), where one can clearly see the change in transport type as $L$ increases, from ballistic to subdiffusive~\citep{Landi2015}.
Finally, in Fig.~\ref{MagneticProfileAnalytic} we compare the magnetization profiles $\langle \sigma_z^i \rangle$ with the exact solutions, which again show perfect agreement. 

\subsection{Regimes with no analytical solution}

\begin{figure}
	\subfloat[]{\includegraphics[width = 3in]{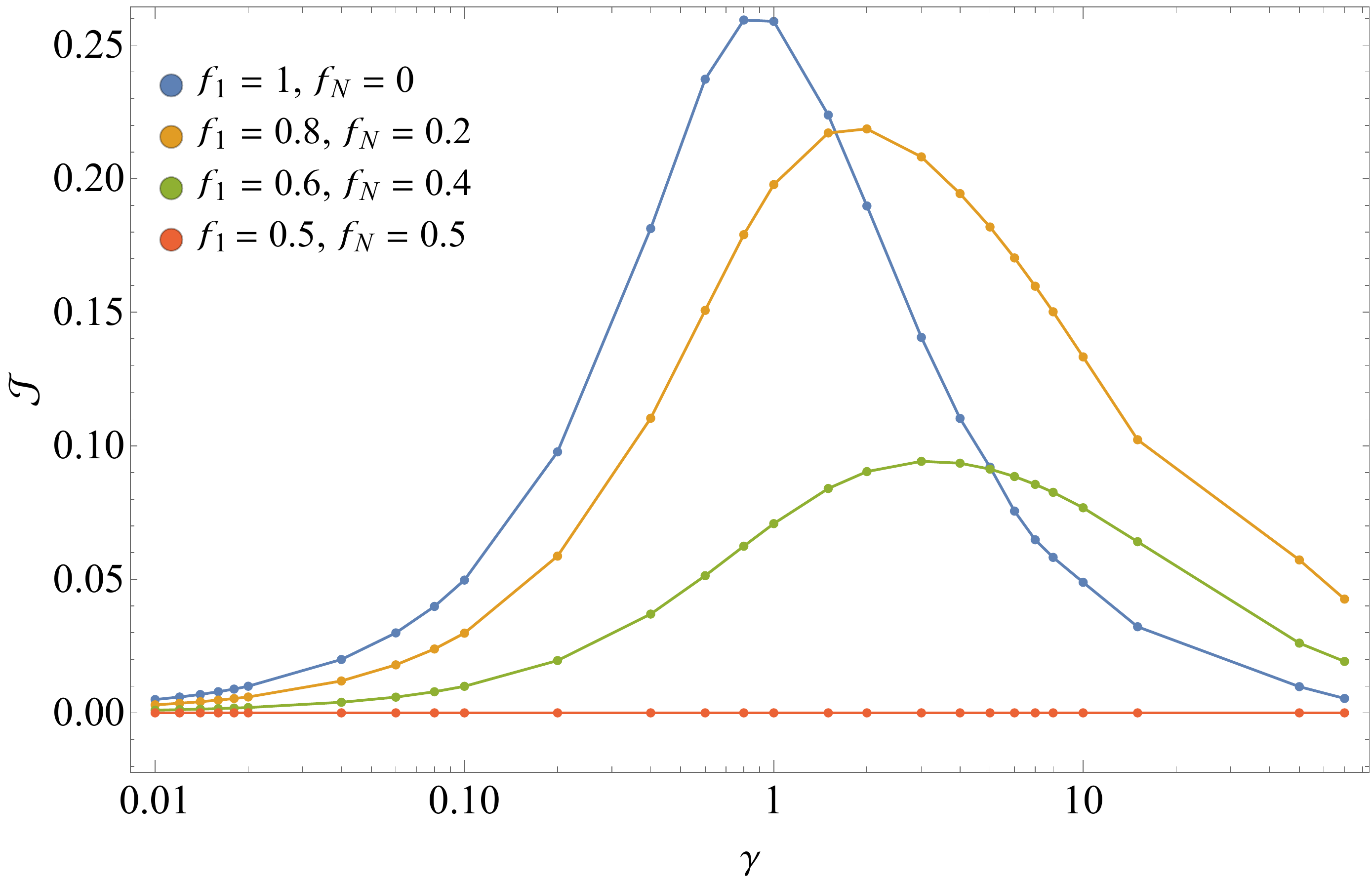}} \qquad 
	\subfloat[]{\includegraphics[width = 3in]{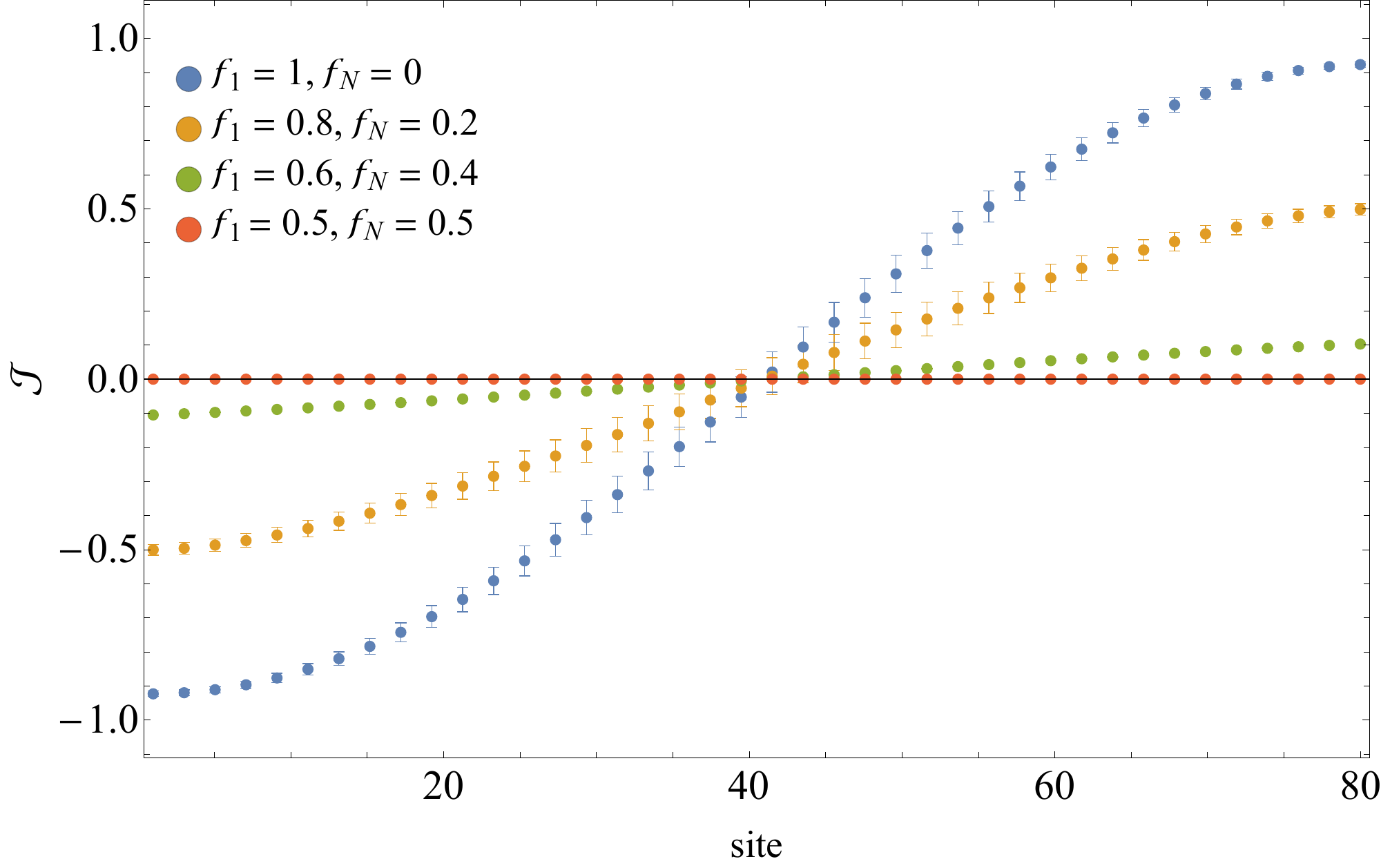}}
	\caption{
	(a) $\mathcal{J}$~vs.~$\gamma$ for $N = 10$ and different values of $f_1, f_N$. 
	(b) Magnetization profile for $N = 80$. 
	Other parameters were $\gamma = 0.5, \Delta = 1, h = 0$.}
	\label{GSConvergence-N10_2}
\end{figure}

Finally, we illustrate how our implementation can be used to explore situations which have no analytical solution and therefore rely {\it solely} on numerical methods. 
The current as a function of $\gamma$ for different driving parameters $f_i$ (see Eqs.(\ref{D},\ref{D_vec})) and $N = 10$ is illustrated in Fig.~\ref{GSConvergence-N10_2}(a).
As can be seen, changing $f_i$ brings significant changes to the steady-state and indeed it vanishes when $f_1 - f_N=0$.  
A similar analysis of the magnetization profile is presented in Fig.~\ref{GSConvergence-N10_2}(b).

%
%
%
%

%
%

%
%

\section{\label{sec:conclusions}Conclusions}

To summarize, in this paper we have detailed an implementation of a DMRG routine suited for dealing with open quantum chains. The implementation is based on the algorithm first presented in \cite{LdLTechnique}, and was implemented on the iTensor library~\cite{ITensor}. The code is also freely available at~\cite{oDMRG}.  
The goal of the implementation is to convert the open problem into the language of traditional DMRG, for which many sophisticated routines have already been developed. 
A major advantage of this method is that it provides, without any overhead, a simple and effective quantifier of convergence, because the steady state corresponds to the zero energy eigenstate of an effective Hamiltonian.  
We have presented several benchmarks, analyzing both the convergence of the algorithm as well as comparing it with  analytical predictions that are available for a limited choice of parameters. 
These analyses clearly show that our implementation is suitable for studying transport properties in one-dimensional quantum chains, and it can thus be used to study quantum transport phenomena such as interaction induced current rectification and negative differential conductance.

\break
\section*{Acknowledgments}
The authors acknowledge fruitful discussion with J. Goold, M. Stoudenmire and L. Gregório. 
G. T. L. acknowledges the financial support of the S\~ao Paulo Funding Agency FAPESP (Grants No. 2017/50304-7, 2017/07973-5 and 2018/12813-0) and the Brazilian funding agency CNPq (Grant No. INCT-IQ 246569/2014-0). 
DP acknowledges support from Ministry of Education of Singapore AcRF MOE Tier-II (project MOE2018-T2-2-142). 
GTL acknowledges the hospitality of apt44, where part of this work was developed.
HPC acknowledges Douglas Casagrande for his support with some of the numerical simulations. 

\appendix
\section{\label{sec:app}Available functions (and how to use them)}

\noindent \verb+LRN-sites.h+
\\

This class is constructed with the $(LR)^N$ formalism in mind, and it therefore rearranges the indices of a N-sized tensor network accordingly. It defines the right $(R)$ and left $(L)$ versions of $S_x$, $S_y$, $S_z$, $S_+$, $S_-$, as well as combinations of those, which are used in the construction of more complex operators, such as $(S_-)R (S_-)L$ and $(S_+ S_-)L$, and so on. It is fitting to be used by any new implementations, and its functionality is akin to the iTensor implemented \verb+SpinHalf+ class.
\\

\noindent \verb+MakeIVEC+
\\

The MakeIVEC function sets the entries of a tensor network to match those of $\text{vec}(I)$. It is called as  
\begin{lstlisting}[frame=single]
    MakeIVEC(MPS &rho, int N)
\end{lstlisting}
The ordering is done in accordance with the indices of the input and the $(LR)^N$ formalism. 
This functions is used as  the initial guess state that is input into the simulation.
This allows one to start with a valid physical state and also ensures that simulations can be re-done by starting from the same state. It is a void function, overwriting the values in \verb+rho+. 
\\

\noindent \verb+LdLXXZConstruct+
\\

This function handles the creation of the Liouvillian MPO. It can receive either constant values of the parameters $\gamma$, $h$ and $\Delta$, or variable ones, which can be input into a vector. The same applies to the temperatures $f_1... f_N$.
It returns an MPO object, and is called using the following line. 
\begin{lstlisting}[frame=single]
    LdLXXZConstruct(SiteSet &sites, double Delta, vector<int> dissipatorsVec,
        vector<double> dissipatorsTempValues, <double> gammaVec, 
        vector<double> hVec)
\end{lstlisting}
\vskip0.3cm

\noindent \verb+WarmUp+
\\

The warm-up routine functions as a simple set of DMRG sweeps with fixed bond-dimension. It has been explained in detail in Sec. \ref{WarmUpSec}.
It is a void type function, which means it simply overwrites the tensor network object. It is called with the following line.
\begin{lstlisting}[frame=single]
    WarmUp(MPS &rho_L, MPO &M2, double convergenceThreshold = 0.001, 
        Args& args = {"Quiet", false}, Args& argsDMRG = {"Quiet", true})
\end{lstlisting}

\bibliographystyle{model1-num-names}
\bibliography{refsSimulation,refsMPS,refsTransport}
\end{document}